# Machine Code Optimization – Improving Executable Object Code

Clinton F. Goss

Westport, CT, USA. Email: clint@goss.com

**ABSTRACT**

This dissertation explores classes of compiler optimization techniques that are applicable late in the compilation process, after all executable code for a program has been linked. I concentrate on techniques which, for various reasons, cannot be applied earlier in the compilation process. In addition to a theoretical treatment of this class of optimization techniques, this dissertation reports on an implementation of these techniques in a production environment. I describe the details of the implementation which allows these techniques to be re-targeted easily and report on improvements gained when optimizing production software.

I begin by demonstrating the need for optimizations at this level in the UNIX programming environment. I then describe a Machine Code Optimizer that improves code in executable task files in that environment. The specific details of certain algorithms are then described: code elimination to remove unreachable code, code distribution to re-order sections of code, operand reduction to convert operands to use more advantageous addressing modes available on the target architecture, and macro compression to collapse common sequences of instructions. I show that the problem of finding optimal solutions for code distribution to be *NP*-Complete and discuss heuristics for practical solutions.

I then describe the implementation of a Machine Code Optimizer containing the code elimination, code distribution, and operand reduction algorithms. This optimizer operates in a production environment and incorporates a machine-independent architecture representation that allows it to be ported across a large class of machines.

I demonstrate the portability of the Machine Code Optimizer to the Motorola MC68000 and the Digital VAX-11 instruction sets. Finally, metrics on the improvements obtained across architectures and across the optimization techniques are provided along with proposed lines of further research. The metrics demonstrate that substantial reductions in code space and more modest improvements in execution speed can be obtained using these techniques.

## Preface

This revised edition of my 1986 Ph.D. dissertation was originally authored on a CDC 6600[1] using TROFF with TBL and EQN. The text for this edition was provided courtesy of The Internet Archive and initially published online in March, 2011. This revised PDF edition was published online August 22, 2013.

## 1. Introduction

The topic of compiler optimization covers a wide range of program analysis methods and program transformations which are applied primarily to improve the speed or space efficiency of a target program. These techniques are typically applied to a representation of the target program which is, to some degree, removed from the program representation executed by the hardware. The representations on which optimization techniques are applied include source-to-source transformations ([Parts 83], [Schn 73]) down to optimizations on assembly code ([Fras 84], [Lower 69], [McKee 65]).

However, in many program development environments, some significant optimization techniques cannot be performed on any program representation prior to the representation executed by the hardware, without general re-design of that environment.

---

[1] The same CDC 6600 that a group from the Transcendental Students took hostage in 1970 in an anti-war protest. Some of the students, possibly members of the Weathermen, attempted to destroy the computer with incendiary devices. However, several staff and faculty, including Peter D. Lax, managed to disable the devices and save the machine.

It is my thesis that the application of optimization techniques at this level is warranted and can be shown to yield a significant decrease in code space and a more modest improvement in execution speed. As a result, this dissertation describes two aspects of this area in parallel. I explore analysis and optimization techniques from a theoretical viewpoint. Some of these are new and some are extensions of techniques which have been applied in other phases of the compilation process. In addition, I report on the implementation of a production quality Machine Code Optimizer. This optimizer represents the first time that these techniques have been brought together in this fashion and at this level. The performance of this optimizer substantiates the expected speed and space improvements on two target architectures.

In this chapter, I demonstrate the need for optimizations at this level and suggest that such optimizations can he carried out despite the lack of auxiliary information which would normally be available to an optimizer. I also survey existing work in closely related areas and outline the remainder of this dissertation.

## *1.1. Background*

As a working example throughout this dissertation, I will consider the compilation and optimization of programs under UNIX and UNIX-like program development environments. These environments will be considered specifically for machines based on architectures such as the Digital Equipment Corporation VAX-11 ([DEC 77]), Motorola MC68000 ([Motor 84a]), and the Texas Instruments TI32000 ([Texas 85]). I will, unless specifically noted, rely only on features of UNIX which are generally available in program development environments. The scope of architectures considered in this research is discussed later in this section.

In source form, a program consists of a number of modules, each containing one or more subprograms (subroutines, functions, etc.). A compiler for the given high level language reads the source code of a single module, possibly translating into one or more internal forms over which optimization techniques are performed, and produces a single object file. This file contains machine code consisting of instructions to be executed by the hardware, data objects which are operated on by the instructions, and other information.

The architectures to be considered here have a memory area consisting of locations with an associated linear ordering. The locations are numbered sequentially by addresses that follow the ordering. Each instruction on these architectures has an opcode which names the operation to be performed and a number of operands which yield values or give the address of data objects or memory locations to be operated on. Each operand requires an area, called an extension word, in the instruction to hold information on the value or machine address which the operand represents.

The bit representations of the instructions and data objects in an object file are identical to those which will appear in memory when the program is executed, except that references to code and data objects in other modules as well as references to absolute addresses in the current object file are not set. Such references are called relocatable references. Any operand of an instruction containing a relocatable reference is called a relocatable operand and the address it references is called the effective address.

Information regarding where relocatable references are and what they refer to is contained in the relocation information in each object file. The location of a relocatable reference, as specified by the relocation information, is called its relocation point.

When all modules of a program are compiled, the object files are supplied to a system linker. The linker produces a task file which can be directly loaded into memory and executed by the hardware. Such a task file contains areas of code, data objects, and optionally, relocation information. Each area, or segment, is formed by catenating the corresponding areas from each object file, in the order they were supplied to the linker, and resolving relocatable references by installing the actual machine address in each reference.

For a particular high level language, it is typical to organize a set of object files that implement the primitives of the language (e.g. `SIN(x)` in BASIC, `Indexed Read` in COBOL, and `printf()` in C) into a library. Such libraries can be given to the linker, which selects only those object files that contain code or data referenced by other modules already linked into the task file.

This general approach reduces the compilation work necessary to effect small changes in a program: only the affected modules need be recompiled. Since linking object files is far faster than re-compiling the whole program from source, this system greatly speeds development of highly modularized programs.

However, this general approach results in a number of inefficiencies in the code in the task files. Furthermore, the optimization techniques that might remove these inefficiencies must be performed after the link phase on the given architecture.

The first inefficiency arises when linking an object file which contains code for several subprograms. If any of the subprograms or data objects in such an object file is referenced, the entire object file is linked into the task file. This situation frequently arises in UNIX environments where large libraries implementing primitives of various types are linked into an application. Furthermore, it cannot be avoided prior to the link phase, except by restructuring the offending object files.

Another inefficiency deals with the use of the instruction set itself. The given architectures all have a set of addressing modes which can be used to represent the semantics of instruction operands. Included in this set are a number of location-relative modes which yield an effective address by giving an offset from the location of the operand or the start of the instruction. Often, the location-relative modes require less space and yield an effective address that is decoded (by the hardware) faster than absolute modes which simply name the effective address. However, the short offset employed limits the effective address to within a specified distance from the operand. This limitation is called the **span** of the mode.

For example, many machines have several addressing modes for operands of branch instructions. Each addressing mode has its own span restrictions. Often, the most general form allows an arbitrary branch target, but is the most expensive in terms of space and execution speed. Shorter and more efficient forms of branch operands compute their target address relative to the memory address of the branch instruction itself, together with an offset from the start of the instruction. However, the offset must be small, allowing only relatively local branch targets.

Often, these location-relative modes cannot be used in UNIX task files due to the method of linking. After an object file is produced, the sizes of instructions do not change; the linker merely fills in resolved addresses in relocatable operands. Again, this allows a fast linker to perform minimal work when code to a single module is changed.

However, relocatable effective addresses which cannot be confined within the span of a location-dependent mode must be implemented using the most general and usually most expensive addressing mode. Under the UNIX scheme, this includes references to code in other modules as well as all references to static data, since the data appears in memory after all code and can be arbitrarily far away from an operand.

Finally, the installation of location-relative modes is itself limited by the order in which object files are linked. Typically, libraries containing code for primitives are linked in at the end of the text segment. Thus, they tend to be far away from the code which uses them. In many high level languages, primitives tend to be the most frequently referenced routines, and locating them at the end of the text segment may significantly reduce an optimizer's ability to install location-relative modes in operands. Such high-use subprograms need to be placed near their references.

Conversely, code generated from the user's source appears at the beginning, far removed from the data segment containing referenced global variables. This code should appear near the end of the text segment, as close to the data segment as possible.

The inefficiencies described thus far are common across a variety of architectures. These generally include machines with a linear address space which provide several interchangeable addressing modes to access this space. Other than the MC68000, VAX-11, and TI32000 mentioned earlier, the Digital PDP-11 ([DEC 75]), Interdata 8/32, IBM 1130, CDC 6600 Peripheral Processor, and the Prime 400 (see [Bell 71] for a general discussion) are in this class. The above remarks do not apply to architectures with a purely segmented architecture such as the Intel 8086 ([Russ 80]).

The generic techniques for handling the inefficiencies described above are applicable across this full class of architectures. However, the implementation of these techniques for a given architecture is highly dependent on the specifics of the instruction set, addressing structure, and memory model of the target machine. A straightforward implementation of these techniques will be riddled with specific references to the architecture. Therefore, it is of special interest to develop, in conjunction with generic techniques for handling these inefficiencies, a technology for instantiating those techniques in an architecture-independent fashion.

### *1.2. Related Work*

The general topic of compiler optimization has received much attention, with the bulk of the work concentrating on transformations applicable to some intermediate representation between source code and executable code. I have borrowed a number of high level concepts and techniques from a number of sources, applying them with greater effectiveness at the machine code level. While the specific work in each area is reviewed in detail in the relevant sections which follow, I outline the major references in each area: A number of the inefficiencies described above can be partially removed by means of techniques applied at a higher program level. These include the implementation of code elimination of various forms, as described in [Aho 77], the handling of span-dependent instructions at the assembly code level in [Szym 78], and the compression of repeated code sequences at the assembly code level in [Fras 84].

Many techniques appear in the literature which deal with low-level constructs, but are not applicable to machine code optimization due to their effectiveness on an intra-module basis. Thus they are more efficiently applied before the machine code level so that work is not repeated on a given module. These techniques include ordering basic blocks to minimize the number of branches [Raman 84], chaining span-dependent jumps [Lever 80], and peephole optimization [McKee 65].

Comparatively little work has been done on optimizations at the machine code level. The works known to the author are those by [Dewar 79a] on compressing an interpretive byte stream and [Rober 79] on distribution of data throughout the code segment as described in Chapter 4.

### *1.3. Organization of the Dissertation*

In response to the inefficiencies described in Section §1.1 and building on the work surveyed in the last Section, a Machine Code Optimizer (MCO) was constructed to make machine code smaller and faster. The remainder of this dissertation deals mainly with the design, implementation, and performance of the MCO. I concentrate on describing those techniques which, for various reasons, cannot be applied before the link phase of compilation.

The next Chapter gives an overview of the organization of the MCO and outlines the design of particular areas.

Chapter 3, Chapter 4, and Chapter 5 expand on the specific techniques for removing the inefficiencies in Section §1.1. References to related work in each area are given as well as the specific algorithms and their analysis.

Chapter 6 describes a number of techniques relating to recognizing and compressing common sequences of code. These are not used in the current MCO for various implementation or efficiency reasons, but the experience gained is of interest to compiler constructors.

Chapter 7 presents statistics on the space and speed improvements gained by the MCO on VAX-11 and MC68000 code for various high level languages. Statistics on the space and time cost of the MCO itself are also presented as well as the effort of re-targeting the MCO from the 68000 to the VAX architecture.

Finally, Chapter 8 reviews the work, summarizes the results obtained, and proposes lines of future research.

## 2. Design of the MCO

The MCO reads an input task file containing executable machine code, data, and relocation information data for a given architecture, applies various techniques for improving machine code, and outputs another task file which is semantically equivalent.

Briefly, the MCO operates in the following sequential phases: The input task file is read and augmented by a set of dynamic data structures which hold information about the instructions and data of the program. These are built up during instruction parsing in which the byte stream containing program code is partitioned into machine instructions and data areas. This list of instructions and data areas is then partitioned into subprograms during text blocking.

The next phase, called operand linking, is responsible for identifying all relocatable operands and determining what they refer to.

The first optimization performed is code elimination in which unreferenced areas of code and subprograms are removed from the dynamic data structures.

Then, code distribution is performed. Sections of code and data are re-ordered to reduce the average distance between instruction operands and the effective addresses they reference. This transformation by itself does not improve the code, but makes the next technique more effective.

Operand reduction converts each instruction operand to use the least expensive addressing mode which can represent the operand on the given architecture. This operates in two sub-phases: `MINIMIZE` contracts all operands to use the least expensive applicable addressing mode and `LENGTHEN` expands minimized operands as necessary to satisfy constraints on the addressing modes.

Finally, the code relocation phase installs changes in the bit patterns of instructions as a result of the improvement techniques applied and writes the output task file.

One of the design goals of the MCO is to ease the onus of re-targeting the MCO to various architectures. Toward this goal, most of the relevant information about the target architecture is kept in a set of static data structures. They describe the details of the instruction set and addressing modes of the target architecture which are needed by the MCO, especially during Operand Reduction. The static data structures allow the MCO to be largely table driven in areas where re-targeting is an issue.

In this chapter, I give a more detailed description of the dynamic data structures and the phases of the MCO I have just outlined. Particular attention is given to how the phases interface and what their effects are on the dynamic data structures. Certain algorithms as well as the static data structures are described and analyzed in later chapters.

## 2.1. Input

Input to the MCO consists of a single task file. This file contains the following areas of information:

**The Header**: A fixed-size structure containing the sizes of the other areas, the program load address where the program is loaded into memory, and the entry point giving the location of the first instruction to be executed.

**The Text Segment**: A byte stream containing the machine code instructions of the program, possibly interspersed with areas of program data. At execution time, the byte stream is loaded into memory (possibly in a virtual fashion) by a system loader, beginning at the program load address specified in the header. The address at which an instruction is loaded into memory is called that instruction's load address.

**The Data Segment**: A byte stream similar to the text segment, but containing only program data. It is loaded by the system loader either directly after the text segment or at some address specified in the header. The rules as to where the data segment may be loaded in relation to the text segment (e.g. at the next 64-byte boundary) vary depending on the environment. The location where the data is loaded into memory is the data load address.

**The Symbol Table**: A list of structures which map symbolic names onto symbol types and machine addresses.

**Relocation Information**: A list of locations in the text and data segments which reference machine addresses. These may be instruction operands which specify the address of an object in the data segment or a pointer in the data segment initialized to point to another piece of data or an instruction. Each such area specified is the size of a pointer on the target architecture.

Except for the last area, the information required by the MCO in a task file is standard in that such information is logically required for a system loader to be able to load a program into memory.

The Relocation Information is optionally provided by the UNIX linker, which links together object files. On some systems, the linker cannot provide this information in the task file. However, the relocation information is simply distilled by the UNIX linker from similar information in each of the object files it links together. This information must be present in some form in object files in order for a linker to assign proper values to pointers. In this case, the MCO can extract and distill it in the same way that the UNIX linker does.

## 2.2. Instruction Parsing and Internal Representations

After opening the input task file and reading the header, the MCO begins parsing the text and data segments to build an internal representation of the program. First, the contents of the text and data segments are read into buffers in memory. Then the MCO creates a list of text and data nodes in memory to hold relevant information about the program.

Each text node describes a single machine instruction and each data node is associated with a single area of contiguous data. The last node on the list is always a data node which represents the data segment.

[Appendix A](#) provides a detailed description of the fields in a text node and what they represent. Figure 2.1 summarizes the fields of a text node using an example of an instruction on the 68000 architecture as they would appear after instruction parsing. In this example, the target address, S, is represented using the absolute long mode.

```
Instruction: 0A00     jsr S
             0A06
             ...
             0C20  S:

Text Node:
 OPC:   o_jsr   -- instruction opcode
 SIZE:  sz_none -- size of object being
                -- operated on
 IADDR: 0A00    -- address of start of
                -- instruction
 FADDR: (NULL)  -- used for operand reduction
 INSTR: 4EB900000C20  -- instruction bytes
 NEXT:          -- text node of instruction
                -- at addr 0A06
 IBYTES: 6 -- Initial # of instruction bytes
 NBYTES: 6 -- Current # of instruction bytes
 REF:    0 -- # of references to this instr
 JSR:    0 -- # of calls to this instr
 OP[0]:     -- operand descriptor for
            -- first operand
  ADDR:   0 -- relocatable address referenced
  TARGET:   -- target text node of operand
  MODE:   am_abs -- current addressing mode
  OFFSET: 2   -- byte offset of operand bytes
  REG:    NULL -- identity of register(s) used
```

**Figure 2.1** Example Text Node

Each data node holds information pertaining to a single contiguous block of data. The data may be in the text or data segments. Data notes have IADDR, FADDR, NEXT, NBYTES, and REF fields which are identical to text nodes. They also have a `DATA` field which serves the same purpose as the `INSTR` field in the text notes.

The dynamic data structures are built up by an instruction parsing routine. This routine is given a pointer to a location in the input text segment and determines the information needed to initialize a single text or data node for the instruction or data area beginning at that location. The instruction parsing routine depends heavily on the architecture and takes a significant portion of the processing time of the MCO.

For the 68000, the logic of parsing instructions is embedded in a large routine (28 pages of C source) which was tightly coded for speed. When re-targeting to the VAX architecture, a data-driven scheme was used. This routine was small (2 pages of C source), developed and debugged quickly, but still runs about as fast as the 68000 version. This was possible due to the greater orthogonality of the VAX instruction set.[2]

### 2.3. Text Blocking

Instruction parsing organizes the text and data nodes into a single linked list, in the order they were read in. This single list is broken down into a two-level data structure during text blocking.

The text and data nodes are partitioned into blocks, each of which is assigned a block node.

Text blocking is performed in a single pass over the code. A pair of text nodes containing an unconditional branch or subprogram return followed by an instruction with its JSR field set constitutes a partition point. At these points, a new block is formed.

Thus, a block is typically one or several subprograms in the text segment where each block is independent and linked only via the block nodes. After text blocking, the two-level data structure is processed by all subsequent algorithms, rather than the initial single list of text nodes. Figure 2.2 demonstrates this transformation. The specific fields of a block node are described in detail in Appendix A.

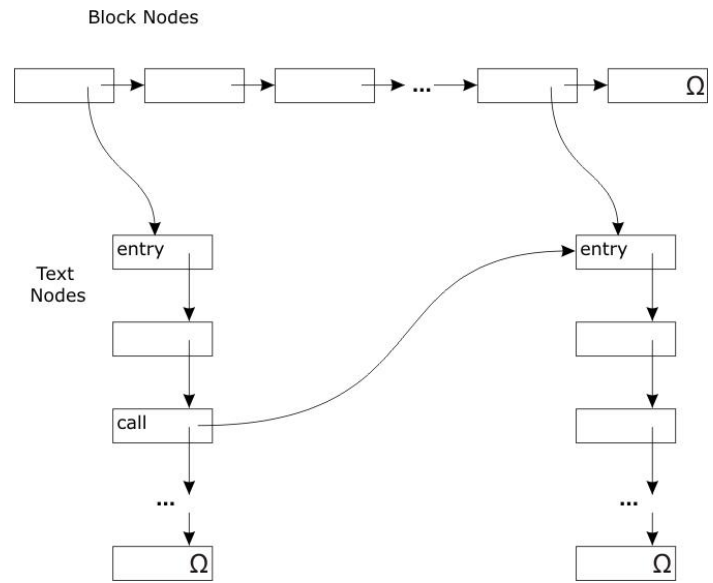

**Figure 2.2**. Text Blocking

Text blocking is done for two reasons. First, the code distribution algorithm reorders sections of code. After text blocking, it simply deals with block nodes rather than lists of text and data nodes. Second, several of the algorithms performed on the dynamic data structures have a worst-case performance which is quadratic in the number of text nodes since they have to perform linear searches for a node with a given `IADDR`.

In these cases, we search through the list of block nodes to find the correct block and then examine the text and data nodes in that block. In this way, the quadratic algorithms run in reasonable time for all but pathological or contrived input.

### 2.4. Operand Linking

After the instructions have been parsed and the dynamic data structures built, the relocatable operands are identified and linked to their targets. This operand linking is done in two passes over the dynamic data structures.

---

[2] One complication of instruction parsing is that no data can appear in the text segment. It is usually straightforward to get the compiler to place constant tables, switch tables, indexed jump tables, etc. into the data segment. However, the VAX implementation was complicated by the presence of register masks at the start of each subprogram. These are arbitrary bit patterns that specify, when control is passed to the subprogram, which registers are to be preserved. If the instruction parser were called with a pointer to a register mask, the resulting text node would be meaningless since the register mask is pure data. Hence, instruction parsing on the VAX cannot be done in a single sequential pass, as it is on the 68000. The VAX implementation runs in two passes. The first pass processes the code sequentially, building a list of known register mask locations from call instructions which are parsed. Any calls to forward targets alert this first pass that register masks exist. However, it is not until the second pass that text nodes are built, when register masks have been marked. This is a source of potential error in the current MCO for the VAX. If the first pass encounters an unmarked register mask, it could mis-parse instructions badly enough to miss another call instruction to a routine which is called only once This routine would then have an unmarked register mask, which would cause problems in the second pass. In practice, the first phase re-synchronizes very quickly (5–10 bytes) and this has not caused problems. For unreferenced subprograms, the instruction parser does attempt to parse register masks during the second pass. However, the text nodes from this parsing will be eliminated during subprogram elimination. A better solution to this problem is to have the compiler or assembler emit a short illegal instruction prior to each register mask. Since execution never flows into a register mask, the marker will do no harm at execution time and can serve as a flag to the instruction parser.

The first pass identifies all relocatable operands. This is done by a pair of co-routines which pass over the text and data nodes and over the relocation information in the input task file. The first co-routine processes instructions up to the next relocation point specified by the second co-routine. The second co-routine processes the relocation information to determine the next text or data node which has a relocatable operand.

The first co-routine marks any operand which uses some location-relative addressing mode as relocatable and the effective address is stored in the ADDR field of the operand.

At a relocation point, we determine whether the address specified as being relocatable is in a text node or a data node. If it is in a text node, the operand of the instruction specified by the relocation information is identified and again marked as relocatable by setting its ADDR field. Since no relocation information is kept in data nodes, relocation directives specifying relocatable references in data nodes are ignored.

Note that, except for location-relative addressing modes, it is crucial to use the relocation information to identify relocatable operands. If we rely on the apparent nature of the operand based on its addressing mode, operands could not be unambiguously identified as relocatable. Consider an instruction which loads the address of its operand.

Although this operand appears to be relocatable, the idioms

```
    lea    val,An    (on the 68000) and
    moval  val,Rn    (on the VAX)
```

are often used to load constant (non-relocatable) values. Conversely, a comparison with an immediate operand may be a constant, but could also be comparing a value with the (relocatable) address of a routine.

After the first operand linking pass, all operands which are relocatable have their ADDR field set. The second pass sets the TARGET field for all such operands. This field is set to point to the text or data node containing the code or data that will be loaded at the ADDR address. Note that the ADDR field need not refer to the start of the code or data in the referenced node; the referenced node must simply contain the target.

It is this second pass which runs in quadratic time in the number of block nodes. However, due to the blocked data structure employed, the second pass runs with reasonable speed (see Section §7.2).

In addition to setting the TARGET field, the REF field of any text node referenced by a relocatable operand anywhere in a text or data node is incremented during pass 2. Also, the JSR field is incremented if the relocatable operand is the operand of some subprogram call instruction.

Thus, at the end of operand linking, the ADDR and TARGET fields are set for all and only those operands which are relocatable. Also, the REF field contains an exact count of the number of relocatable references to a text or data node. The JSR field contains a lower bound on the number of call instructions which refer to a given text node. Due to indirect calls through pointers to procedures, some text nodes which may be the target of a call instruction at execution time cannot be identified. However, such nodes will always have a non-zero REF field.

Figure 2.3 depicts the text nodes from Figure 2.1 after the operand linking phase.

```
Instruction: 0A00      jsr S
             0A06
             ...
             0C20  S:

Text Node:
 OPC:   o_jsr
 SIZE:  sz_none
 IADDR: 0A00
 FADDR: (NULL)
 INSTR: 4EB900000C20
 NEXT:       -- text node of instruction at
             -- addr 0A06
 IBYTES: 6
 NBYTES: 6
 REF:    0
 JSR:    0      TN: Text Node at operand target:
 OP[0]:
   ADDR:   0C20    OPC: ??
   TARGET: TN      IADDR: 0C20
   MODE:   am_abs ...
   OFFSET: 2       REF: 1  -- # references
   REG:    NULL    JSR: 1  -- # of calls
```

**Figure 2.3** Example Text Node After Operand Linking

### 2.5. Code Elimination

The first code improvement performed by the MCO is the elimination of code which can never be executed. As previously described, we wish to eliminate sections of code as well as entire subprograms which are never referenced.

The code elimination algorithm is an augmented version of unreferenced code elimination ([Aho 77]). It relies on the REF and JSR fields set during the previous phase to determine what code can safely be eliminated. This is done in a single forward pass over the code. The algorithm removes unreferenced code as well as certain referenced sections which are not reachable from the program entry.

The algorithm and the restrictions on what input programs it operates on are given in Chapter 3.

### 2.6. Code Distribution

As mentioned in the introduction, one of the inefficiencies in the way UNIX task images are linked arises from the order in which subprograms in the text segments are arranged. Both the order of subprograms in a module and the order in which modules are supplied to the linker give no consideration to placing span-dependent operands near their targets.

The code distribution phase re-orders subprograms in the text segment to place span-dependent operands near their targets. The target might be in another subprogram or in the data segment at the end of the program.

In Chapter 4, I discuss the general problems of re-ordering sections of the data segment as well as the text segment from a theoretical viewpoint and show these problems to be difficult to solve. Due to these results, I employ efficient heuristics to distribute the blocks. These heuristics are described in Chapter 4, along with other related optimization techniques not employed in the current MCO.

### 2.7. Operand Reduction

After code elimination and distribution, the operand reduction algorithm is employed. This algorithm makes aggressive use of the addressing modes available on the target architecture to transform existing instruction operands to make them smaller and faster.

The general operand reduction algorithm is loosely based on the one proposed by [Szym 78] for assembling code for architectures with span-dependent instructions. The correctness and termination arguments in that paper apply in a similar fashion to operand reduction.

As with the assembler algorithm of [Szym 78], we perform operand reduction in two phases. The first phase, `MINIMIZE`, makes a single pass over the code. For instructions and operands which can potentially be reduced (all relocatable operands and certain others with specified addressing modes), we form the set of all legal opcode/addressing mode pairs which can yield a semantically equivalent instruction. We then choose the shortest combination and install it.

After `MINIMIZE`, the `LENGTHEN` phase iterates over the code identifying operands which employ addressing modes that are unsuitable due to some span-dependent constraint on the mode. Again, we form the set of possible opcode/addressing mode substitutions. We now choose the least expensive one which satisfies all semantic as well as span-dependent constraints.

Again, no change is made to instruction bit patterns, but sufficient space is maintained in the text node to store the full instruction.

The operand reduction algorithm is implemented in a largely machine-independent fashion using static data structures to describe the necessary attributes of the instruction set and addressing modes of the target architecture. The details of these static data structures and the associated algorithms outlined above are given in Chapter 5.

### 2.8. Code Relocation

The final phase of the MCO, code relocation, installs the changes made during earlier code improvements in the bit patterns of each instruction and data area and produces an output task file.

First, a single pass is made over the code to install new bit patterns in instruction opcodes and operands which were subject to operand reduction. The lengths of instructions are correctly maintained by `MINIMIZE` and `LENGTHEN`, so no re-allocation of buffers to hold instruction bytes is needed during code relocation.

Next, the relocation information in the input task file is re-scanned to find any relocation directives referring to relocatable addresses in the data nodes. We then modify the pointer value in the referenced data node to contain the `FADDR` of the node whose `IADDR` was equal to the input pointer value. If the input pointer referenced an instruction, we make sure that it points to the start of the instruction. An input pointer to the middle of a data area can safely be translated since data nodes are never contracted or expanded. We simply add the same offset to the `FADDR` address that the original pointer was offset from the `IADDR` of the target data node.

Finally, a file header for the output task file is written, followed by the contents of each text and data node. To satisfy the requirements of UNIX debuggers, a copy of the input symbol table modified to reflect the changes in the machine address for each symbol is also output.

### 3. Code Elimination

The code elimination phase of the MCO removes subprograms which can never be invoked. It uses an augmented version of a simple single-pass code elimination algorithm that employs a good heuristic to test which sections of the flow graph have circularities but are not connected to the program entry point. In this way, entire subprograms, especially ones with loops, can be eliminated in a single pass.

This chapter begins by describing the limitations to which the input program must be subject in order for this technique to be applicable. I then give a classification of existing code elimination techniques and present my algorithm in light of these.

*3.1. Restrictions on the Input Program*

In general, the problem of code elimination on machine code is complicated by two considerations: First, since we are not improving code from a given high-level source language, we cannot rely on any rules of program structure (e.g. a task image compiled from PASCAL source would never have a jump into the middle of a subprogram). Rather, we must accept any valid machine code generated from any high-level language.

This problem is handled by the MCO by using the relocation information to set the REF field of all instructions whose IADDR is referenced by some operand. Hence, each instruction is treated separately and no assumptions regarding program structure need be made.

The second problem arises since arbitrary machine code can appear. For example, it is possible to compute a jump address from non-relocatable operands without having the computed address named in the relocation information.

The problem of identifying these situations is undecidable, since the expression that computes a referenced address may be arbitrarily complex and take arbitrary inputs.

Consider the following pseudo-machine code:

```
load  addr(X), reg1
add   17, reg1
jump  *reg1
```

In this scheme, the node containing *X* will have its REF field set since the load instruction has a relocatable reference to it. However, the node at *X – 17* will not be marked as referenced, and could be erroneously subject to code elimination.

Hence, I restrict the target of all (direct and indirect) control transfers to conform to the following definition: A branch address is simple if it is identical to the initial address of some operand or data area specified as relocatable by the relocation information.

This requires that the compiler generate only simple branch addresses in order for the REF field of nodes to be accurate.

Note that the definition of a simple branch target does **not** rule out constructs such as:

**Jumping to an address which was extracted from an array**. Such code is typically generated for the BASIC ON-GOTO, Fortran Computed-GOTO, and C switch statements. Since each entry in the array referring to an address is specified as a relocatable data area, all possible targets of such high-level statements will be marked as referenced.

**Pointers to code and procedure parameters**. Such code appears in C procedure pointers and Fortran ASSIGN statements. The pointer values are generally loaded from a data area, as above, or by some code such as

```
load  addr(X), reg
```

in which case the first operand of the load will be marked as relocatable.

**Interrupts and service routines**. Although these routines are called asynchronously in response to some event, their address appears somewhere in the task file.

Typically, some interrupt vector or table needs to be initialized when the program begins. This is done either by installing the address of the routine in a table using executable code (the address would then appear as a relocatable operand of an instruction), passing the address of the routine to a system function (again, the parameter passing mechanism would contain the relocatable operand), or by initializing the table directly in the data segment (a relocatable data item would be in the data segment).

Hence, I do not feel that the requirement for simple branch operands is a practical restriction on compilers for most high-level languages. In fact, the MCO has been used with production compilers for full ANSI COBOL ([Phil 85co]), FORTRAN 77 ([Phil 85ft]), C ([Phil 85c]), and two versions of BASIC ([Phil 84cb], [Phil 84mb]). None of the code generators or any of the library code for the language primitives had to be modified to accommodate this restriction.

*3.2. Current Code Elimination Techniques*

I define several methods for performing code elimination, with increasing degrees of effectiveness:

**Unlabeled code elimination** removes code which follows an unconditional branch and is not labeled. This can be done using a single forward pass over the code.

**Unreferenced code elimination** eliminates code following an unconditional branch which is either unlabeled or is prefaced with a label which is not referred to. After reference counts are tabulated on labels, unreferenced code is eliminated by a series of converging forward passes over the code.

**Unreachable code elimination** eliminates code to which there is no flow path from the entry point of the program. It is capable of eliminating, for example, mutually recursive subprograms whereas unlabeled and unreferenced code elimination are not. Typically, this technique is implemented by building a flow graph for the program and removing disconnected subgraphs which do not contain the entry point.

### 3.3. Subprogram Elimination

For purposes of the MCO, unlabeled code elimination is not effective since every instruction is potentially labeled.

Unreachable code elimination, although the most aggressive technique, is also not applicable. In the presence of an indirect jump through a quantity in a register, the MCO would need to trace all possible values in the register to determine the possible successors to a flow graph node. Failing this, all nodes would need to be labeled as successors, rendering the entire flow graph connected. Such jumps often arise when generating code for high level constructs such as C `switch` statements and COBOL `perform` statements, so this technique would yield poor results.

Due to the presence of reliable reference counts, unreferenced code elimination is most suitable for the MCO. However, it fails to fully eliminate a subprogram that has a loop in it. Consider the following code:

```
            return
    SUB1:   instr
            instr
    LOOP:   instr
            instr
            jump    LOOP
            instr
            return
    SUB2:   instr
```

where `SUB1` is an unreferenced instruction, `LOOP` is referenced due to the loop, and `SUB2` is the beginning of the next subprogram which is referenced.

Basic unreferenced code elimination removes code from `SUB1` up to, but not including, `LOOP` (denoted `SUB1~LOOP`). However, `LOOP` up to `SUB2` (`LOOP~SUB2`) is not eliminated since `LOOP` is referenced.

In order to eliminate such routines, I implement an augmented version of unreferenced code elimination called **subprogram elimination**. This eliminates likely sections of code on a trial basis and checks the resulting program for consistency. When a candidate for unreferenced code elimination is detected, we perform the following algorithm:

1. Let `SUB1` be an unreferenced instruction following an unconditional branch, return, etc. Let `LOOP` be the first referenced instruction following `SUB1` and let `SUB2` be the first instruction with its `JSR` field set at or after `LOOP`. Perform basic unreferenced code elimination and remove `SUB1~LOOP`.

2. Decrement the reference count of any instruction which is the target of an operand of an instruction in `LOOP~SUB2`.

3. Scan `LOOP~SUB2` and determine if any instructions are still referenced.

4. If no instructions in `LOOP~SUB2` are referenced, then any instructions in `LOOP~SUB2` which were referenced before step 2 were the target of operands within `LOOP~SUB2`. The instructions in `LOOP~SUB2` can he eliminated and further code elimination continue at `SUB2`.

5. Otherwise, some operand outside the range LOOP~SUB2 has the referenced instruction found in Step 3 as a target. The instructions in LOOP~SUB2 cannot be eliminated and we must repair the damage done to the reference counts in Step 2.

Further code elimination proceeds from the jump to `LOOP`.

Whenever an instruction is removed during the subprogram elimination algorithm, we decrement the reference count of any targets of operands of the instruction. (This is handled differently in steps 2, 4, and 5 above, but the net effect is the same). If the resulting reference count goes to zero and the target is not currently being eliminated (precedes `LOOP` in the above algorithm), another opportunity for code elimination has occurred. However, since subprogram elimination works by forward passes only, this opportunity will not be caught on this forward pass. Hence, we repeat subprogram elimination until no such situations arise.

### 4. Code Distribution

The code distribution phase of the MCO re-orders sections of a program to improve the effectiveness of operand reduction.

I divide the task of re-ordering a program into two problems:

**Problem 4.1 (Data Distribution)**

> Partition the data segment into independent data objects, each of which can be moved without regard for the load location of other independent data objects. Then reallocate these objects in slots in the code segment which are not on an execution path (e.g. following a return or unconditional branch) in order to place them closer to operands which reference them. ¤

**Problem 4.2 (Code Distribution)**

> Re-order the code blocks created during text blocking to reduce the distance between inter-block branches and their targets. ¤

This division corresponds to improving the effectiveness of operand reduction as it deals with two distinct types of operands; those which reference targets in the data segment and those which reference other code blocks.

In this chapter, I examine both these problems as implemented in [Rober 79] and the MCO, respectively.

### *4.1. Data Distribution*

Given the class of code improvements to which the MCO is addressed (those which can only be done at or after the link phase), an algorithm for data distribution would be appropriate in the MCO. However, a number of problems prevent it from being implemented in this application.

The first problem is the lack of a reliable way of partitioning the data segment into independent data objects, which preserves the semantics of the input program. The input data segment is seen as a single block of data. No information is provided regarding what areas of the data segment must remain in a fixed position relative to other areas. For example, it is unclear where one array ends and another begins.

To obtain a complete partitioning, information as to the layout of the data segment would have to be provided by the compiler for each module in the task file. This is feasible, but is outside the current design of the MCO.

The second problem with data distribution concerns the ramifications of placing modifiable program data in the same area in memory as program code or constant data.

All code currently generated by the compilers with which the MCO operates is reentrant, thus allowing the text segment to be shared in a multi-task environment. Data distribution renders the code non-reentrant since data would be interleaved in the text segment. Thus it could not be used where text is shared or on a system where the text segment is protected by hardware support.

Another problem with data distribution concerns the expectation, on the part of the programmer, that the data segment will be laid out in the order that static data is declared in the source code. Although the layout of static data is usually unspecified in language standards, compilers have had no reason to allocate data in other than the input order.

As a result, the folklore for certain languages dictates that certain programming constructs which rely on the order of static data are acceptable.

For example, a well-known technique in FORTRAN for building a zero-based array of integers is to declare as follows:

```
INTEGER   DUMMY
INTEGER   A(99)
```

where `DUMMY` becomes an alias for `A(0)`. Although illegal, this usage is not detectable in general, not flagged as an error even in specific cases where it is detectable (e.g. constant subscript), and actually works on all FORTRAN compilers known to the author (FTN [Contr 75], FORTRAN System/370 [IBM 74], Philon FAST/FORTRAN [Phil 85ft]).

Finally, there are the issues of actually performing data distribution in reasonable time and space. The problem of data distribution was first examined in [Rober 79] from a theoretical viewpoint. He showed that the problem of finding an optimal solution to Problem 4.1 is *NP*-Complete [Garey 79]. Furthermore, the problem of finding a solution which is within a (non-trivial) constant factor of the optimal solution is also *NP*-Complete.

Thus, the best we could hope for is a well-tuned heuristic which places variables well. In the UNIX environment, where data resides at the end of the text segment, even a simple heuristic could improve the code substantially. For example, one might go through the independent data objects in order, placing each in the slot which maximizes the number of references to it which can be made short at the time. In the absence of the problems already mentioned (e.g. on a single-user dedicated machine with no memory protection), such a heuristic might be worthwhile.

### *4.2. Complexity of Code Distribution*

The problem of code distribution as stated at the beginning of this Chapter is characterized in graph theoretic terms as follows:

> A **directed graph** *G = (V, E)* consists of a finite set *V* of **vertices** and a finite collection of edges, *E: V × V*, where each edge connects a pair of distinct vertices in *V*. The collection of edges of a graph may have duplicates (**parallel edges**) but the set of vertices may not. Edges are denoted $\mu \rightarrow v$ where $\mu, v \in V$. If $\mu \to v \in E$ then vertex $\mu$ is adjacent to $v$. The set of all vertices adjacent to vertex $v$ is denoted *adj(v)*. A weighted graph $W = (V, E, W_v, W_e)$ is a directed graph with functions $W_v: V \to (N = \{0, 1, ...\})$ and $W_e: E \to N \times N \times N \times N$.

In characterizing code distribution, we map the code blocks of the text segment onto the vertices of a weighted graph and use the edges to represent the inter-block references.

We allow parallel edges since a block may reference another block many times, but we allow no self-loops (edges must connect distinct vertices) since intra-block references are not considered.

The weight function on vertices, $W_v$, gives the size, in bytes, of the code block as read in from the text segment. The weight function on edges is formulated from the position of the source and destination of the inter-block reference within their respective blocks. In Figure 4.1, the source of the reference is offset $s$ bytes in code block $\alpha$ and the target is at position $t$ in code block $\gamma$. Thus, $W_e(\alpha \to \gamma) = (s, s', t, t')$.

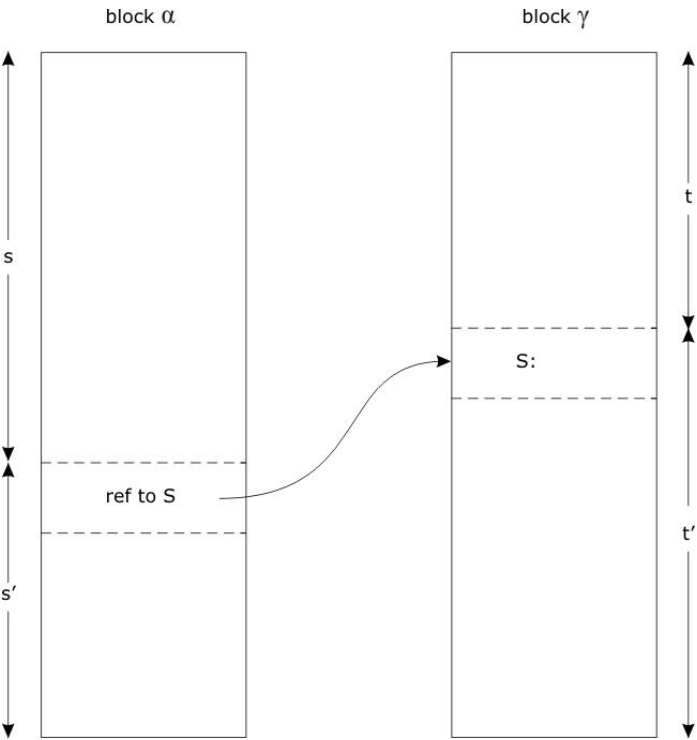

**Figure 4.1**. Inter-block Reference

The goal of code distribution is to find a permutation of the vertices, $\psi: V \leftrightarrow \{1, 2, ..., |V|\}$, which keeps the number of edges requiring a long addressing mode to a minimum. Given a permutation, $\psi$, for each edge $\mu \to v$, we define:

$span(\mu \to v) = endpoints(\mu \to v) + interposed(\mu \to v)$

$endpoints(\mu \to v) =$ **if** $\psi(\mu) < \psi(v)$ **then**

　　$W_e(\mu \to v)(2) + W_e(\mu \to v)(3)$

**else**

　　$W_e(\mu \to v)(1) + W_e(\mu \to v)(4)$

$$interposed(\mu \to v) = \sum_{\substack{i > \min(\psi(\mu), \psi(v)) \\ i < \max(\psi(\mu), \psi(v))}} W_v(\psi(i))$$

The span of an inter-block reference must account for the location of the source and destination of the reference in their respective blocks (the *endpoints()* function) as well as the size of all intervening blocks in the ordering of code blocks (the *interposed()* function).

Given a weighted graph $W = (V, E, W_v, W_e)$, a permutation $\psi: V \leftrightarrow \{1, 2, ..., |V|\}$, and a threshold $T$, we define the threshold cost function:

$TCF(W, \psi, T) = |\mu \to v \in E : span(\mu \to v) \geq T|$

The problem of code distribution is analogous to the problem `MINLTA`:

Problem 4.3 (`MINLTA` - Minimum Linear Threshold Arrangement)

> Given a weighted graph $W = (V, E, W_v, W_e)$, we wish to find a permutation $\psi: V \leftrightarrow \{1, 2, ..., |V|\}$ which orders the vertices such that the threshold cost function, $TCF(W, \psi, T)$, for a given threshold $T$, is minimized. ¤

`MINLTA` relates to code distribution as follows: We wish to order the code blocks in the text segment to minimize the number of inter-block references whose span exceeds a certain threshold $T$.

For example, in Figure 4.2, if blocks $\alpha$ and $\gamma$ are ordered with block $\beta$ between them, then the span of $\alpha \to \gamma$ is the sum of $s'$, $r$, and $t$. In `MINLTA`, the $s'$ and $t$ are incorporated in $endpoints(\alpha \to \gamma)$ while $r$ is represented in $interposed(\alpha \to \gamma)$.

I now show that the decision version of `MINLTA` is *NP*-Complete [Garey 79] by

1. showing that the problem can be solved non-deterministically in polynomial time and
2. by polynomially reducing instances of a related problem, `MINLA`, to instances of `MINLTA` such that `MINLTA` yields the same answer as `MINLA` would have.

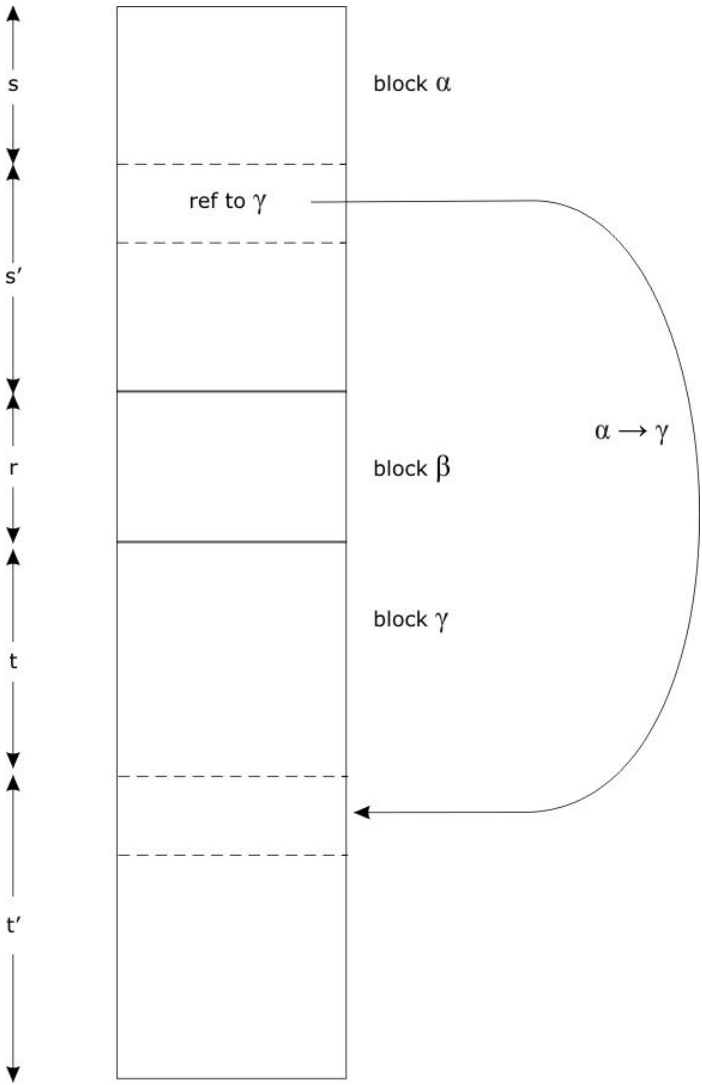

**Figure 4.2**. Computing the Span of a Reference

This cost function is identical to the *span($\mu \to v$)* function given for MINLTA with $W_v(v) = 1$ and $W_e(\mu \to v) = (0, 0, 0, 0)$ for all vertices and edges, respectively. ¤

A simpler version of Problem 4.5, in which the graph was undirected, was shown to be *NP*-Complete in [Even 75] and [Even 79] by a two-stage reduction from the maximum cut set problem on graphs.

Theorem 4.1

The decision version of MINLTA is *NP*-Complete.

Proof: First, we assert that MINLTA can be solved non-deterministically in polynomial time. This is done by non-deterministically choosing the appropriate permutation, $\Pi$, from the $O(|V|!)$ permutations of the vertices and evaluating *TCF(W, $\Pi$, T)*.

Next, we reduce instances of MINLA to instances of MINLTA: Given an instance of MINLA consisting of $G = (V', E')$ and an integer $k$, we define an instance of Problem 4.4 as follows: The vertices $V$ of $W$ are the same as those of $V'$ of $G$. For each edge $e \in E'$, $E$ contains a **bundle** of $|V|-1$ edges $e_1, e_2, \ldots, e_{|V|-1}$. The weight of an edge $W_e(e_i) = (i, i, i, i)$. The weight of all vertices $W_v(v) = 2$ (see Figure 4.3).

Problem 4.4 (Decision Version of MINLTA)

Given a weighted graph $W = (V, E, W_v, W_e)$, a threshold $T$, and an integer $k$, is there a permutation $\psi: V \leftrightarrow 1, 2, \ldots, |V|$ which orders the vertices such that *TCF(W, $\psi$, T) $\leq k$*. ¤

Problem 4.5 (MINLA - Minimum Linear Arrangement)

Given a directed graph $G = (V, E)$ and a positive integer k, is there a permutation $\psi: V \leftrightarrow 1, 2, \ldots, |V|$ which orders the vertices such that the **additive cost function**, *ACF(G, $\psi$) $\leq k$*, where:

$$ACF(G, \psi) = \sum_{\mu \to v \in E} |\psi(\mu) - \psi(v)|$$

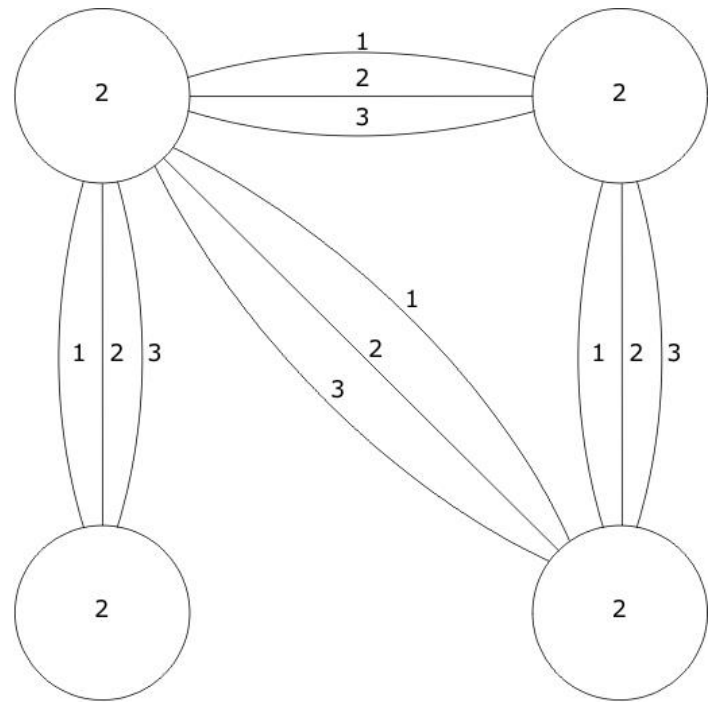

**Figure 4.3**. Mapping MINLA onto MINLTA

I propose that any ordering function, $p$, on $V$ will yield the same value for *TCF(W, Π, 2|V|-2)* under MINLTA as *ACF(G, Π)* under MINLA.

**Case 1**

Consider two vertices, $\alpha$ and $\beta$, which are placed sequentially by $\Pi$. Under MINLA, an edge $\alpha \rightarrow \beta$ between them contributes one to *ACF(G, Π)*. Under MINLTA, exactly one of the edges between $\alpha$ and $\beta$ from among those generated from $\alpha \rightarrow \beta$ yields a *span ≥ 2|V|-2* (the edge with weight *(|V|−1, |V|−1, |V|−1, |V|−1))* so one is added to *TCF(W, Π, 2|V|−2)*.

**Case 2**

Under MINLA, if there are $n$ vertices interposed between the endpoints of $\alpha \rightarrow \beta$, then the edge $\alpha \rightarrow \beta$ adds $n-1$ to *ACF(G, Π)*. Likewise, under MINLTA, exactly $n-1$ edges from the bundle of edges generated from $\alpha \rightarrow \beta$ would be included in *TCF(W, Π, 2|V|−2)*. Those are the edges with weights *(|V|−n−1, |V|−n−1, |V|−n−1, |V|−n−1)* through *(|V|−1, |V|−1, |V|−1, |V|−1)*. Hence, an ordering, $\Pi$, of $V'$ yields *ACF(G, Π) ≤ k* if and only if that ordering of $V$ yields *TCF(W, Π, 2|V|−2) ≤ k*.

*Q.E.D.*

In the light of these results and the expectation that the number of code blocks in a text segment is on the order of the number of subprograms, an algorithm for code distribution which yields an optimal solution is not likely to run in polynomial or reasonable time on a deterministic processor. However, it should be noted that there are some differences between MINLTA and the problem of code distribution:

Since current architectures often have location-relative modes using byte, word, and long offsets, the real-world code distribution problem could be required to deal with several thresholds rather than just one.

In code distribution, the size of a code block is not fixed, but depends on the placement function itself. MINLTA simplifies this by assigning a span value for an edge which is determined solely from the initial conditions of the problem, while the operand reduction algorithm must be applied for each placement function to determine the span of an edge.

In MINLTA, we allow the weights on edges to be arbitrary positive numbers, while in practice they would be limited to the weights of their corresponding vertices. It is not known whether this more restrictive version of MINLTA is *NP-Complete*.

### 4.3. Heuristics for Code Distribution

Since the possibilities for an efficient optimal algorithm for code distribution are dim, the MCO applies a heuristic to order the code blocks.

The basic approach is to build a tuple of code blocks starting at the end nearest the data segment. At each step, we choose the best block from among those yet to be placed, according to a heuristic which evaluates unplaced blocks. This block is added to the start of the tuple. This basic scheme is summarized in the algorithm:

```
proc basic_code_distribution();

  unplaced := {set of blocks};
  set_of_spans := {spans for addressing modes
                    of this architecture};
  placed : = [];

  while unplaced ≠ {} do
    bestworth := -1;
    (∀ bl ∈ unplaced)
      w := 0;
      (∀ span ∈ set_of_spans)
        w -:= worth(bl, unplaced, placed,
                                  span);
      end ∀;
      if w > bestworth then
        bestworth := w;
        bestbl := bl;
      end if;
    end ∀;

    placed := [bl] + placed;
    unplaced less:= bl;
  end while;
end proc;
```

Of course, the effectiveness of this algorithm depends on the `worth(bl, unplaced, placed, span)` function. The MCO currently uses two heuristic functions in combination:

$\sigma_0$    This function evaluates references in `bl` to the data segment. The following multiplicative factors constitute $\sigma_0$:

$\sigma_0^1$    The number of references in `bl` to the data segment.

$\sigma_0^2$    The fraction of the data segment that the average reference (i.e. one at the center of `bl`) would reach under the given span if `bl` were placed at the head of the placed list.

$\sigma_0^3$    The number of bytes saved by installing the addressing mode associated with span over the addressing mode with no span restrictions.

$\sigma_0^4$    The inverse of the size of `bl` (larger blocks are penalized). This may be thought of in combination

with $\sigma_0^1$ to produce a single factor which denotes the density of data references per byte of code.

$\sigma_1$    This function evaluates references between `bl` and blocks already placed in the list. The following factors constitute $\sigma_1$:

$\sigma_1^1$    The number of code references in `bl` which would reach their targets under the given span in the placed list if `bl` were placed at the head of the list.

$\sigma_1^2$    The number of code references in the list which would reach the average target in `bl` under the given span.

These functions are designed to choose heuristically what would seem to be the best block from among the remaining unplaced blocks when running the inner loop of `basic_code_distribution()`. The $\sigma_0$ function accounts for expected gains from operand reduction due to references to data. Likewise, $\sigma_1$, predicts gains from references to the code already in the list. Within each of these functions, $\sigma_0^1$, through $\sigma_0^4$, and $\sigma_1^1$ and $\sigma_1^2$ can be balanced to give the best results.

These functions are implemented efficiently by attaching the following information to each block node, `bl`:

**REF**

For each span, the number of references to text nodes in `bl` from blocks in the placed list which reach `bl` under the span.

**RELOC**

The number of relocatable operands in text nodes in `bl` which reach nodes in block in the placed list under each span.

**DRELOC**

The number of relocatable operands in text nodes in `bl` which reach nodes in the data segment.

These fields are maintained by the following expanded algorithm:

```
proc code_distribution();

  unplaced := (set of blocks};
  $ Set the DRELOC field of each block.
  set_dreloc(unplaced);
  placed := [];
  plsize := 0;

  while unplaced ≠ {} do
    bestworth := -1;
    (∀ bl ∈ unplaced)
      $ Modify REF and RELOC to account for
      $ the most recent
      $ block added to the list and
      $ references which are
      $ now out of range.

      update_ties(bl, placed);
      w := 0;
      (∀ span ∈ set_of_spans)
        w -:= worth(bl, unplaced, placed,
                            span, plsize);
      end ∀;

      if w > bestworth then
        bestworth := w;
        bestbl := bl;
      end if;
    end ∀;

    placed := [bl] + placed;
    plsize +:= size(bl);
    unplaced less:= bl;
  end while;
end proc;
```

The MCO allows any combination of $\sigma_0$ and $\sigma_1$, to be used during a run. The relative effectiveness of these heuristics is reported in Chapter 6.

## 5. Operand Reduction

As described in Section §2.7, operand reduction installs, in each operand, the least expensive addressing mode which satisfies all constraints imposed by the architecture.

This Chapter begins by describing the data structures which represent the attributes of the target architecture needed for operand reduction. This is followed by a discussion and analysis of the algorithms which implement the two phases of operand reduction.

### 5.1. Static Data Structures

At the heart of the `MINIMIZE` and `LENGTHEN` phases, the following problem arises:

Problem 5.1. (Build Translation Class)

> Given an instruction, *i*, and an operand of that instruction, op, form the set of (opcode, addressing mode) pairs which can be used in place of the existing opcode of *i* and addressing mode of *op (OPC(i), MODE(op))*. This set is called the `TRANSLATE_CLASS(i, op)`. ¤

The remainder of this section describes how the `TRANSLATE_CLASS` is built.

First I describe what types of restrictions the target architecture places on membership in this set. Then I give a set-theoretic description of how the `TRANSLATE_CLASS` set is formed. Finally, I discuss a space and speed efficient implementation of the set formers.

The set-formers and algorithms in this Chapter are presented in the set-theoretic language SETL ([Dewar 79b]) to elucidate the concepts involved. Lower level versions, coded in C ([Kern 78]), may be found in Appendices B and C.

For an opcode/addressing mode pair `(opc, am)` to belong to the set `TRANSLATE_CLASS(i, op)`, it must satisfy the following restrictions:

**Addressing Restrictions**: Under the rules of the target architecture, am must be a legal addressing mode for an operand of opcode `opc` in operand position `OPNUM(op)`. Furthermore, the new opcode must accept the same number of operands as the existing opcode and, for each operand, `op'`, of `i` other than the operand being considered, `MODE(op')` must also be legal for the new opcode in that operand position.

**Semantic Restrictions**: Each addressing mode on the given architecture performs a function such as yielding a value of some type or operating on a register. The function of am must be equivalent to that of `MODE(op)`. Likewise, the function of the new opcode, `opc`, must be equivalent to the existing opcode, `OPC(i)`.

**Span Restrictions**: If `am` is a location-relative mode, then the effective address which op yields must be within the span of am.

To form a `TRANSLATE_CLASS` which complies with the addressing and semantic restrictions, we begin with the following sets defined across all opcodes, `opc`, and addressing modes, `am`:

ADDRESSING_CLASS(opc, opnum)

> For each operand position, opnum, corresponding to an operand of `opc`, the set of addressing modes which are legal on the target architecture.

OPERAND_EQUIV_CLASS(am)

> The set of addressing modes which perform an equivalent function to `am`.

OPCODE_EQUIV_CLASS(opc)

> The set of opcodes which perform an operation which is equivalent to `opc`.

For a given instruction, `i`, and operand, `op`, the `TRANSLATE_CLASS(i, op)` is formed, as needed, by the following set constructors:

$$OPERAND\_TRANSLATE\_CLASS(i, op) = OPERAND\_EQUIV\_CLASS(MODE(op)) \cap \left( \bigcup_{opc \,\epsilon\, OPCODE\_EQUIV\_CLASS(OPC(i))} ADDRESSING\_CLASS(opc, OPNUM(op)) \right)$$

For the given instruction and operand of that instruction, this is the set of addressing restrictions of the machine for that instruction and operand and the semantic restrictions imposed by the existing addressing mode.

$$OPCODE\_TRANSLATE\_CLASS(i, op) = \{ opc \,\epsilon\, OPCODE\_EQUIV\_CLASS(OPC(i)) \mid$$
$$(ADDRESSING\_CLASS(opc, OPNUM(op)) \cap OPERAND\_TRANSLATE\_CLASS(i, op) \neq \Omega) \text{ and}$$
$$(\forall (opnum \neq OPNUM(op))\ MODE(op[opnum]) \,\epsilon\, ADDRESSING\_CLASS(opc, opnum)) \}$$

This is the set of opcodes which are equivalent to the current opcode and which allow at least one addressing mode in the `OPERAND_TRANSLATE_CLASS()` for the given opcode. Also, the current addressing mode in operands we are not scrutinizing must be allowed in the operand position of each opcode in this set.

$$TRANSLATE\_CLASS(i, op) = \{ (opc, am): opc \,\epsilon\, OPCODE\_TRANSLATE\_CLASS(i, op),$$
$$am \,\epsilon\, (OPERAND\_TRANSLATE\_CLASS(i, op) \cap ADDRESSING\_CLASS(opc, OPNUM(op))) \}$$

Finally, we combine the intermediate sets to form the `TRANSLATE_CLASS()` as defined above, satisfying addressing and semantic restrictions, but not span restrictions.

In practice, we do not form the OPERAND_TRANSLATE_CLASS and `OPCODE_TRANSLATE_CLASS` sets, but construct `TRANSLATE_CLASS` directly.

The following algorithm presents a high level view of how `FORM_TC` is implemented:

```
proc FORM_TC(i, op)
  TRANSLATE.CLASS := {};
  (∀ opc ∈ OPCODE_EQUIV_CLASS(OPC(i)))
    if opc ≠ OPC(i) then
      $ Check that the operands of opc other
      $ than op accept the current
      $ addressing modes in i.

      (∀ opnum ∈ [1..NOPER(opc)] |
                    opnum ≠ OPNUM(op))
        if MODE(opnum) ∉
              ADDRESSING_CLASS(opc, opnum) then
          continue opc;
        end if;
      end ∀;
```

```
      end if;

      (∀ am ∈ ADDRESSING_CLASS(opc, OPNU.VI(op)))

         $ Check that this new mode is
         $ semantically equivalent to the
         $ existing mode.

         if am ∈ OPERAND_EQUIV_CLASS(MODE(op))
                                               then
            TRANSLATE_CLASS with:= [opc, am];
         end if;
      end ∀;
   end ∀;
end proc;
```

I now represent the data structures and algorithms for `FORM_TC` in a lower level implementation. The data structures were designed to conserve space and be accessible with reasonable speed. The version of `FORM_TC` as coded in C is presented in Appendix C.

In order to represent the `ADDRESSING_CLASS`, `OPERAND_EQUIV_CLASS`, and `OPCODE_EQUIV_CLASS` sets, a set of static data structures are built for the given architecture. The static data structures consist of a pair of tables, one for addressing modes and one for opcodes, and various arrays as described below.

First, we examine the addressing mode table. This is an array of addressing mode descriptors, one for each distinct addressing mode on the target architecture. Two addressing modes in two different instructions are considered distinct if they are represented differently in the two instructions or are not semantically equivalent. In particular, modes which are represented using different bit patterns or the same pattern in different locations in instructions must be distinct.

Consider the Data Register Direct addressing mode on the 68000 `mov` instruction. As a source operand, this mode is the same for `mov` as for the first operand of a `cmp` instruction. However, a distinct addressing mode must be used for a destination operand of `mov` which uses Data Register Direct since the location of the bits to specify the mode and register are in a different location in the instruction.

Each addressing mode descriptor contains the following fields which are relevant to this discussion:

**SIZE, SPEED**

> Values used to evaluate the cost of using an addressing mode. These are relative values used for purposes of evaluating cost functions and are related to the clock cycles and size in bytes above a basic opcode for the use of the addressing mode.

**OEC**

> Pointer to an array of nodes, each containing the code of an addressing mode in the same `OPERAND_EQUIV_CLASS` of this mode. All modes in the same OEC have the same effect on the relevant aspects of the machine state when evaluated.

**SPAN_OK**

> A pointer to a predicate which determines, given an instruction and an operand of the instruction, whether the addressing mode would satisfy span restrictions if installed.

**INSTALL**

> A pointer to a routine to install the addressing mode in a given instruction and operand. This routine is invoked during code relocation.

The opcode table contains a single opcode descriptor for each distinct opcode on the target architecture. As with addressing modes, a single operator is sometimes broken down into several opcodes for purposes of operand reduction even though the bit patterns of the instructions may be identical. This occurs in multi-operand operators since the addressing mode in the `ADDRESSING_CLASS(opc, opnum)` must all be valid regardless of addressing modes employed in other operands.

Operators such as the 68000 `sub` instruction must be broken down into two opcodes: a `sub_d` opcode which allows a large class (source class) of addressing modes as a first operand and a data register for a second operand and a `sub_m` opcode whose first operand is a data register and whose second operand can be represented using another set of addressing modes (memory alterable class). To implement these using a single opcode would imply that the sub instruction allows any source class mode and any memory alterable mode in its two operands, which is not the case.

An opcode descriptor contains the following relevant fields:

**NOPER**

> Number of operands accepted by this instruction.

**SPEED**

> Used in evaluating opcode/addressing mode pairs. This is the speed relative to other instructions in the operand equivalence class.

**OPEC**

> An opcode which is in the same `OPCODE_EQUIV_CLASS` as this opcode. The `OPEC` fields of all opcodes in a non-singleton `OPCODE_EQUIV_CLASS` set form a circular linked list using this field.

**CLASS**

> An array of pointers, one for each operand of the opcode. Each pointer names an array of nodes containing the codes of addressing modes in the `ADDRESSING_CLASS` of this opcode and operand.

The structure of these tables is summarized in Figure 5.1. I show an example of the static data structures of two instructions on the Motorola 68000: `jmp` and `bra`. These instructions are semantically equivalent, so their `OPEC` fields form a ring. However, the sets of addressing modes allowed for their respective operands are disjoint. Each of the addressing modes is described by an addressing mode descriptor. Finally, the semantic meanings of addressing modes are related in the operand equivalence classes. Through this data structure, a `jmp` using the absolute long addressing mode can be converted to a bra using the disp8 mode.

## 5.2. Minimize and Lengthen

The purpose of operand reduction is to find an optimal solution to the following problem:

Problem 5.2. (Operand Reduction)

> Install the least expensive addressing mode in each operand of each instruction so that all addressing, semantic, and span restrictions are satisfied.¤

In the last section, I presented a general algorithm to find all opcode/addressing mode substitutes for a given instruction and operand that satisfy addressing and semantic constraints. The remaining problem of operand reduction is to satisfy span constraints.

This problem is examined in [Rich 71] and [Fried 76]. In [Szym 78], two algorithms are presented which produce optimal solutions. I will briefly describe the requirements and complexity of each before presenting my solution.

The first, which I call Algorithm *Sz1*, builds a graph to represent the program.

Each operand of each instruction which can employ a location-relative mode is represented by a node in the graph.[3] A directed arc *A→B* is installed if the instruction for *B* lies between *A* and a target which references an operand of *A* in the program. In each node, information similar to our own text node is maintained. In addition, for each operand, the distance from the instruction to the target of the operand (the operand's range) is maintained.

All operands represented by nodes are initially assigned a minimum length location-relative addressing mode. We then process nodes in the graph whose range exceeds the span of the current addressing mode. A longer addressing mode with a larger span is then installed and all predecessors of such nodes in the graph (i.e. nodes whose range depends on the size of the expanded instruction) have their ranges increased to accommodate the longer addressing mode. The node may then be removed from the graph if a maximum-length addressing mode has been installed. The algorithm terminates when no more nodes need to be expanded.

Algorithm *Sz1* produces an optimal assignment of addressing modes using a graph with $O(n)$ nodes and $O(n^2)$ arcs. [Szym 78] claims that the running time, with suitable low-level data structures, is at worst $O(n)$ since each node must be visited at most once for each addressing mode.

In practice. Algorithm *Sz1* is useful for the application described in [Szym 78] jump or subprogram call operands on the Digital Equipment Corporation PDP-11 ([DEC 75]).

Under this instruction set, a single location-relative addressing mode whose span is approximately ±256 bytes is available for such operands. This limits the out-degree of nodes in the dependency graph to 255 for contrived pathological cases. In practice, the average out-degree is 3.5 (across a large sample of application code) which allows *Sz1* to operate rapidly in practical cases.

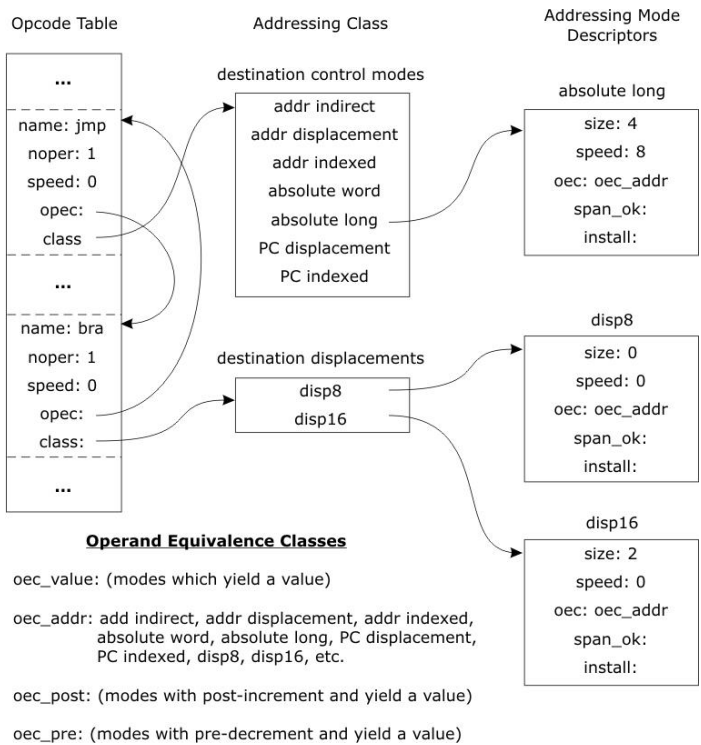

**Figure 5.1**. Static Data Structures for 68000 `jmp` and `bra`

[3] Since Szymanski applied his technique to assembly language before it was assembled, he only considered operands of branch and subprogram call instructions.

However, each of the target architectures, in addition to the PDP-11 mode described above, has location-relative modes with spans of approximately ±32,767 bytes. This allows the out-degree of nodes to be at most 16,381 (assuming a minimum of two bytes per instruction) and an average of 896 in practice. These figures render *Sz1* impractical for our use, especially since we wish to process not only branch operands, but all relocatable operands.

Algorithm *Sz2* is similar to *Sz1*, except that the arcs are not represented in the graph. Instead of adjusting the range of predecessors in the graph, whenever an operand is expanded, a brute-force scan of the instructions is made to find operands whose range need adjustment. This reduces the space requirements to $O(n)$ but the running speed goes to $O(n^2)$.

Again, since the maximum span of an addressing mode is ±254 bytes on the PDP-11, only a small area of code needs to be scanned when an operand is expanded.

However, for this application, the re-scanning often requires a large portion of the program, thus rendering the running time quadratic in practice.

My algorithm builds on *Sz2* with the same worst-case space and time complexity, but runs in linear time in practice. Rather than maintaining the range of an operand, the range value is computed as necessary. This can be done since the `TARGET` field has been set for all such operands during operand linking.

As with *Sz1* and *Sz2*, the operand reduction algorithm begins with `MINIMIZE`, which performs a single pass over the code. For each instruction, `i`, and relocatable operand, `op`, we change the opcode and addressing mode to the pair from `TRANSLATE_CLASS(i, op)` which yields the shortest instruction:

```
proc MINIMIZE()

  (∀ b ∈ BLOCK_LIST)

    (∀ tx ∈ TEXT(b) | tx is a text node)

      (∀ op ∈ OP(tx))

        tc := FOR.M_TC(tx, op);
        bestcost := MAXCOST;

        (∀ [opc, am] ∈ tc)

          c := cost(tx, op, opc, am);

          if c < bestcost then
             bestcost := c;
             newpair := [opc, am];
          end if;
        end ∀;

        if newpair ≠ [OPC(tx), MODE(op)] then
           contract(tx, op, newpair);
        end if;
      end ∀;
    end ∀;
  end ∀;
end proc;
```

After `MINIMIZE`, the `LENGTHEN` phase installs larger addressing mode in operands using a series of passes over the code. The first step in each pass is to set the `FADDR` field of each text and data node to reflect its current load location based on the sizes of all instructions before it. This is the field we will later use to determine the ranges for operands.

We then process each relocatable operand of each text node. If a location-relative addressing mode, am, is currently in use, the range of the operand is computed using the `FADDR` field of the instruction and the `FADDR` field of the `TARGET` node of the operand. The predicate `SPAN_OK(am)` is then evaluated for the range to determine if the operand needs expansion. If so, we compute the `TRANSLATE_CLASS(instruction, operand)`. From this we choose the least-cost opcode/addressing mode pair for which `SPAN_OK(am)`, evaluated for the range, indicates that the new mode satisfies all span restrictions.

This phase is summarized in the following algorithm:

```
proc LENGTHEN()

  change := true;
  while (change) do

    change := false;
    $ Set FADDR fields of all nodes.

    addr := IADDR(TEXT(BLOCK_LIST(1))(1));
    (∀ b ∈ BLOCK_LIST)
      (∀ tx ∈ b)
        FADDR(tx) := addr;
        addr += NBYTES(tx);
      end ∀;
    end ∀;

    $ Expand operands as necessary.

    (∀ b ∈ BLOCK_LIST))
```

```
     (∀ tx ∈ b | tx is a text node)

       $ Get all relocatable operands
       $ (ones with TARGET set) which
       $ might need expansion.

       (∀ op ∈ OP(tx) | TARGET(op) ≠ Ω and
               loc_relative(MODE(op)))

         range := FADDR(TARGET(op)) –
                    FADDR(tx);
         if SPAN_OK(MODE(op))(range) then
           continue;
         end if;

         tc := FORM_TC(tx, op);
         bestcost := MAXCOST;

         (∀ [opc, am] ∈ tc)

           c := cost(tx, op, opc, am);

           if  c < bestcost then
             bestcost := c;
             newpair := [opc, am];
           end  if ;
         end ∀;

         if [OPC(tx), MODE(op)] * newpair then
           expand(tx, op, newpair);
           ehange := true;
         end if;
       end ∀;
     end ∀;
   end ∀;
 end while;
end proc;
```

This algorithm performs well in practice since range values are changed only at the start of each pass and are done through the TARGET pointer rather than maintaining explicit range values in operand descriptors. The TARGET field generally requires $O(n^2)$ time to compute but, during operand linking, we compute these efficiently using the blocked dynamic data structure (see Section §2.3). Likewise, $O(n)$ passes could be made through the code during LENGTHEN, giving an $O(n^2)$ worst case. In practice, the algorithm converges in 2–5 iterations (see Section §7.2).

### 5.3. Register Tracking

The design of operand reduction, as described thus far, falls short in one major area: it utilizes index modes which use only the program counter, while many architectures allow indexing off other registers. For example, on the Motorola 68000, even if a target is not within the span of a PC-indexed mode, if an address register points in the vicinity of the target, an address register-indexed mode is available which costs the same space and time as the PC-indexed mode.

Hence, an improvement to the current operand reduction algorithm would be to provide a data structure which maintains the known values in all registers which can be indexed. In addition, known values in non-indexable registers may be useful since such registers can replace addressing modes which yield constant values.

A number of approaches can be taken in handling this data structure:

1. Have the compiler set aside a single address register as a base register, thus mimicking segmented architectures such as the Intel 8086. This register could be initialized by the MCO to point to an advantageous location and references to all targets which fall within the span of this location could be improved. This is essentially the scheme taken in the Macintosh operating system ([Apple 85]) for the Macintosh 68000-based computer. However, on the Macintosh, all data references must be made using the base register; this limits global data to 32,767 bytes on this machine.

2. Allocate base registers on a less global level. Information as to which registers are unused over ranges in the code would have to be obtained. These could then be initialized and used as local base registers if there were sufficient references in the range which could index off the address register.

3. Information as to which registers have known values in ranges of the code could be obtained by techniques similar to constant propagation [Aho 77]. These registers could be used as base registers in the proper ranges without initialization. On architectures such as the Motorola MC68020 [Motor 84b] where a number of registers can be combined with scaling factors and constant offsets, registers could be used in linear combinations to produce the least expensive addressing mode.

### 6. Macro Compression

Until this point, I have described optimizations and techniques that are employed in the MCO and that are, to varying degrees, successful toward the goals of optimizing task files and furthering this research. In this Chapter I reflect upon a class of techniques that are also consonant with this research but which did not yield satisfactory results in some dimension of performance and were removed from the production version of the MCO.

## 6.1. Background

Common code compression is a class of optimization techniques in which common sequences of code are identified by various analysis methods and removed by altering the code or providing information to a translator that is converting the code to a lower level.

This class of techniques includes common subexpression elimination, available expression elimination, very busy expression hoisting, and code hoisting and sinking (see [Aho 86] for a general discussion of these). These techniques are generally more suitable to earlier phases of the compilation process than the link phase.

The technique of macro compression recognizes common code sequences and replaces each occurrence of the common code with a call to a code macro or subprogram containing the common code.

This space optimization was first used in [Dewar 79a] to conserve space in an interpretive byte stream. The language used an 8-bit opcode but only had 80 operators.

The remaining 176 opcodes were used to represent frequently occurring byte sequences beginning on instruction boundaries. In practice, only multi-bytes instructions or part instructions were subject to macro compression, but the savings remained substantial.

The theoretical aspects of this problem were studied in [Golum 80]. The assumptions were:

1. A byte stream was to be minimized;
2. A macro call consisted of a single byte;
3. Exactly *m* macros of *length* ≤ *k* were to be chosen.

Optimal polynomial-time solutions were obtained which characterized potential macro choices within the byte sequence using an interval or overlap graph (depending on two slight variations of the problem). However, these algorithms were very costly in practice.

## 6.2. Assembly Code Compression

A more recent approach [Fras 84] has been to apply pattern matching techniques to assembly code to identify repeated subsequences. A suffix tree ([McCr 76]) is built for the input code to be compressed. The suffix tree for a list of instructions, *i*, is a tree whose |*i*| leaf nodes are labeled with the locations in *i* and whose arcs are labeled with subsequences of *i*. For example, if *a*, *b*, and *c* are instructions and *$* is the unique end marker, the instruction list *abcab$* would have the tree shown in Figure 6.1.

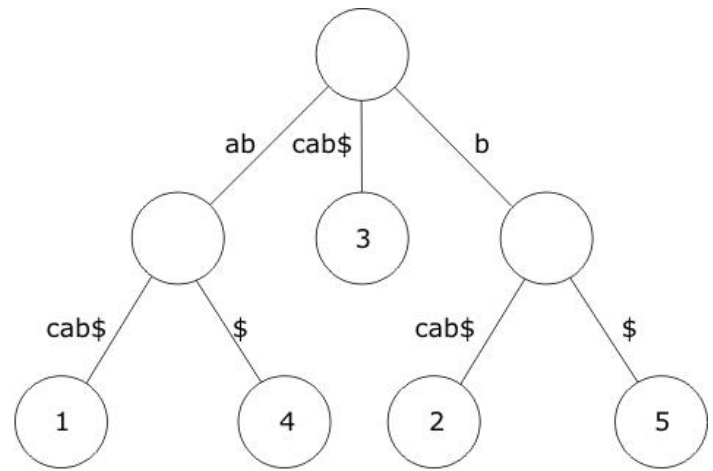

**Figure 6.1**. Example Suffix Tree

This data structure allows us to find the subsequence beginning at any position and ending at $ by following the path of edges from the root to the leaf with the proper label.

More importantly for macro compression, each non-leaf (internal) node represents a common subsequence: the text of the subsequence is found by following the edges from the root to the internal node and the number and location of the subsequences are represented by all leaves whose path to the root goes through the internal node.

Once the suffix tree is built (in linear time — see [McCr 76]) the internal nodes of the suffix tree are evaluated for validity under the semantic rules of macro compression for the given assembly language and for payoff if they were replaced. Valid subsequences are ordered in a priority queue by some criterion and the items of the queue are processed in order, installing a code macro and calls to it at each step.

An optimizer for assembly code was built by [Fras 84] and was reported to run efficiently and perform well. However, no statistics were given on the amount of compression achieved.

## 6.3. A First Attempt

A preliminary optimizer for assembly language was built along these lines for the purpose of gathering statistics. As expected, the effectiveness of macro compression heavily depended on the size of the assembly code file. In assembly files generated from languages such as COBOL ([ANSI 74]), a good deal of compression was obtained since the entire user program is generated in a single assembly code file. However, for languages such as C where a high degree of modularity tends to be observed, almost no compression was obtained. Furthermore, code for language primitives, since they are relatively small and selectively linked modules, were never compressed.

The next stage was to build an analyzer which maintained statistics on common sequences across assembly language files. If, for a given language and compiler, many common sequences appeared repeatedly across different programs, a database of those sequences could be made available to a peephole optimizer [McKee 65]. It would replace them in linear time [Knuth 77] and with small space overhead. A call to a macro body would be substituted on the expectation that the sequence would appear enough times in the various modules of the program to make substitution worthwhile, on the average.

The macro bodies would then be selectively linked in from a large library of these subprograms.

However, it was found that, while a single program may have many common subsequences within itself, the same sequences were, for the most part, not shared between programs. Table 6.1 gives a summary of common sequences in two test programs we will describe in detail in Chapter 7. These programs are called `p1.68` and `p2.68` when compiled for the 68000. A breakdown is given for various sized subsequences for `p1.68`, `p2.68`, and sequences which appeared in both. In each case, I report the number of common sequences as well as the average number of occurrences of each sequence in the programs. In the last column, I report the average occurrences in both programs combined.

These figures show that even though the same compiler was used and the same code for language primitives was linked in, few sequences were common to both program in comparison to the program treated separately. This happens since many of the common sequences contain code which refers to program-specific global data or subprograms.

**Table 6.1** Tabulation of Common Sequences

|       | p1.68 |       | p2.68 |       | p1.vx |       |
|-------|-------|-------|-------|-------|-------|-------|
| Size  | #     | occur | #     | occur | #     | occur |
| 50–60 | 1     | 2     | 2     | 2     | 0     |       |
| 40–48 | 1     | 2     | 2     | 2     | 0     |       |
| 30–38 | 3     | 2     | 2     | 2     | 0     |       |
| 20–28 | 4     | 2.25  | 4     | 3.25  | 0     |       |
| 18    | 3     | 2.00  | 4     | 7.00  | 0     |       |
| 16    | 6     | 3.33  | 4     | 2.75  | 0     |       |
| 14    | 9     | 4.00  | 7     | 5.71  | 1     | 4.00  |
| 12    | 12    | 3.17  | 13    | 6.07  | 2     | 4.50  |
| 10    | 17    | 6.65  | 13    | 7.30  | 0     |       |
| 8     | 32    | 6.00  | 48    | 7.96  | 10    | 17.80 |
| 6     | 41    | 13.46 | 61    | 16.84 | 14    | 52.29 |
| Total | 129   |       | 160   |       | 27    |       |

### 6.4. Macro Compression in the MCO

The results of the assembly code macro compressor indicated that a macro compressor which operated on all the modules in a single program would compress the most macros. This, the macro compressor was recoded to operate on task files, and this became the first version of the MCO. It operated with essentially the same instruction parser and code relocation algorithm described earlier, but without any other optimizations described so far.

There were a number of significant additions to the MCO implementation beyond that of [Fras 84]:

1. The user had the choice of two priorities when inserting sequences into the priority queue: they could be inserted in order of the number of bytes saved by the substitution of the sequence (assuming no overlap with earlier substitutions) or they could be inserted based on the number of bytes in the sequence ('value' priority versus 'length' priority).

2. The MCO was more aggressive in salvaging sequences which would have been discarded as invalid: if the sequence referenced the stack, the macro body was constructed so that the return location was not stored on the stack: if the sequence modified the stack in certain simple ways, similar transformations were applied to the macro body; If some code in the sequence would have caused the sequence to be discarded and that code appeared near the end of the sequence, it was shortened and re-inserted into the priority queue.

3. Since [Fras 84] does not describe the data structures in detail, it is not clear how their suffix tree was represented internally. The MCO maintained the tree in virtual memory. The initial implementation had a pointer from each internal node to the first child and a pointer from each node to its sibling. This simple data structure was very compute intensive during the construction of the suffix tree (48:07 for `p1.68`) and a hash table was installed to represent the parent-child relation between the root node and the second level of the tree (reducing the time to 3:22 for `p1.68`).

The statistics relating to this version of the MCO are reported in Tables 7.2 for the execution time of macro compression, 7.5 for the size improvement in the text segment, and 7.6 for the degradation in the target program's execution speed. From these results it was decided that macro compression was not desirable in the production version of the MCO because of the costs in the following areas:

**Compressor Speed**

The macro compressor required several times the compute resources of the other optimizations combined. The total speed of about 125 bytes/sec was very close to the speed of the [Fras 84] implementation, but we were dealing with the entire task file on each run of the optimizer.

**Dynamic Memory Requirements**

The suffix tree nearly doubled the dynamic memory required by the MCO.

**Execution Speed Degradation**

When running a program after macro compression, the task image is smaller, but the CPU must spend time executing the macro calls. On p1.68, the compressed program took 15% more CPU time than the original (with macro compression and operand reduction). In real time, the compressed file took 7% more time - probably due to the fact that the operating system treats smaller task files with a higher priority. This speed degradation is not as severe if the length priority is used rather than value priority (see Section §7.5). With value priority 913 calls to macros were installed whereas for length priority 821 calls to macros were installed.

Thus, fewer macro calls are made and execution speed is not affected as much with length priority. In addition, value priority saved a total of 2,310 bytes while length priority saved 2,384 bytes.

*6.5. G-Code and G-Compression*

G-compression takes the concept of macro compression to extremes. First, the text segment is converted into a very compact **generative code** or G-code. At execution time this is loaded into memory along with the data segment for the program, a decoder, and 68 an execution buffer. The decoder is responsible for re-constituting sections of G-code into their original native code and placing them in the execution buffer. Code is executed in the execution buffer until a new section of code needs to be re-constituted, at which time control returns to the decoder. If enough space is allocated for the execution buffer and a good allocation algorithm is used, the decoder will be called infrequently compared to execution of native code in the buffer. Even if the processor spends half its time in the decoder, this is substantially better than the 10 to 40 times speed degradation experienced in typical interpretive systems, with the potential for a greater savings of space.

This approach is similar in concept to 'throw-away compiling'. This technique compiles frequently interpreted sections of code at run-time ([Brown 76], [Brown 79], [Hans 74]). However, the task of code generation was employed at run-time in these systems to produce object code, rather than a straightforward decoding. Hence, the translation was slow, required a very large 'decoder', and could not achieve a high level of object code optimization.

To elaborate on these ideas, I describe the items in memory at execution time in more detail:

**G-Code**

The generative code is a representation of the text segment of the original program in which a series of transformations have been applied to translate original instruction and sequences of instructions into G-code instructions. The first set of transformations modifies certain types of instruction operands. In general, references to registers or registers with displacements are unaffected. However, references to instructions or data in the original code are converted to references to the corresponding instructions or data in the G-code.

Each reference to an instruction in the original text segment (original text reference) is replaced by the bit address of the start of the corresponding (j-code instruction relative to the start of the sequence of G-code instructions (soft text reference). Thus the number of bits to represent a soft text reference depends on the size of the G-code.

Each reference to a byte in the original data segment (original data reference) is replaced by the byte offset of the referenced data byte from the start of the data segment (soft data reference). The number of bits for a soft data reference depends on the size of the data segment.

Any immediate operand which refers to the address of an instruction in the original text segment is replaced by the soft text address of the corresponding G-code instruction.

An immediate operand which refers to an original data item is converted to a byte offset from the start of the data segment.

In general, the set of operand addressing modes defined for native code and G-code differ in order to accommodate these transformations. Also, since there are no byte-boundary limitations imposed by the decoder, operands can occupy any number of bits and can even vary in size. Native code designs incorporating some of these features such as bit-aligned instructions and variable sized operands have been developed in the Intel 432 architecture ([Tyner 81]) and the design of the Burroughs B1700 ([Wiln 72]).

The second transformation applied to the original text segment is solely for the purpose of compressing the instruction sequence. The criteria are that it must be decodable starting at any instruction boundary and that the decoding must be done in real time (requiring no look-ahead)

[Peter 61]. This can be accomplished by one or a combination of:

1. A straightforward Huffman encoding ([Knuth 73], p. 402) of the G-code.
2. A partial Huffman encoding in which the main instruction word is encoded but words which contain addresses and displacements are not affected.
3. A macro compression scheme in which common code sequences are collapsed into a macro table (which becomes part of the G-code) and replaced with short non-instructions from a Huffman encoding list. First, the priority queue is ordered by the number of times a sequence appears, rather than by the length or value of a sequence. Then all sequences which appear more than once are inserted into the queue. When processing the queue, sequences are replaced by bit encodings of increasing length. These encodings are assigned in the way that Huffman codes are built [Gall 78] so that minimum space is required.

**Data Segment**

This is identical to the data segment which would be loaded with the original version of the program, except that it is not necessarily loaded at the same address. Pointers in the data segment to other data addresses are relocated based on the new base address for the data segment. Text addresses are translated to their corresponding soft address.

**Decoder and Execution Buffer**

The decoder is a fixed section of code which runs on the native hardware. Its job is to re-constitute G-code into machine code. It takes sections of G-code which needs to be executed and decodes them into a variable-sized execution buffer. The decoded native code is essentially the same as the original native code in the text segment. The differences are that addresses which refer to the data segment are adjusted to point to the new data segment and text addresses are converted back from soft addresses into the hard addresses of the decoded section, if the code at the target text address is already in the execution buffer. Otherwise, a branch to a non-decoded code section consists of a push of the soft address and a call back to the decoder.

The decoder has the following entry points:

MAIN: This is the entry point from outside the program. The first section of code is re-constituted, the user's stack, registers, and arguments are initialized, and the first section of code is called.

TRANSFER: Branch to this entry point to re-constitute and execute code beginning at the soft address which is on top of the stack. All branches to TRANSFER which are preceded by a push of this soft address are then converted to a branch directly to the newly re-constituted native code.

EXTEND: Append a new section of native code after the last executed block. Two items are on the stack: the soft address of the new code and the hard address to begin placing the hard-code. Accessed from unconditional branches in the original program code which are at the end of a re-constituted block, this is really a special case of TRANSFER which can optimize speed by eliminating unconditional branches.

CALL: Same as TRANSFER, except that a subprogram call has been made.

The contents of memory during program execution is summarized in Figure 6.2.

Aside from a straight executable program, the G-code scheme can be used in ways more closely tied to the machine. For example, G-code might be the actual language of the machine, while the decoder resides in microcode itself. The macro bodies themselves would be read in when the program is loaded. The rudiments of such a scheme are employed in the VAX-11 architecture ([DEC 77]), which allows the microcode for certain instructions to he read in by the user during system initialization.

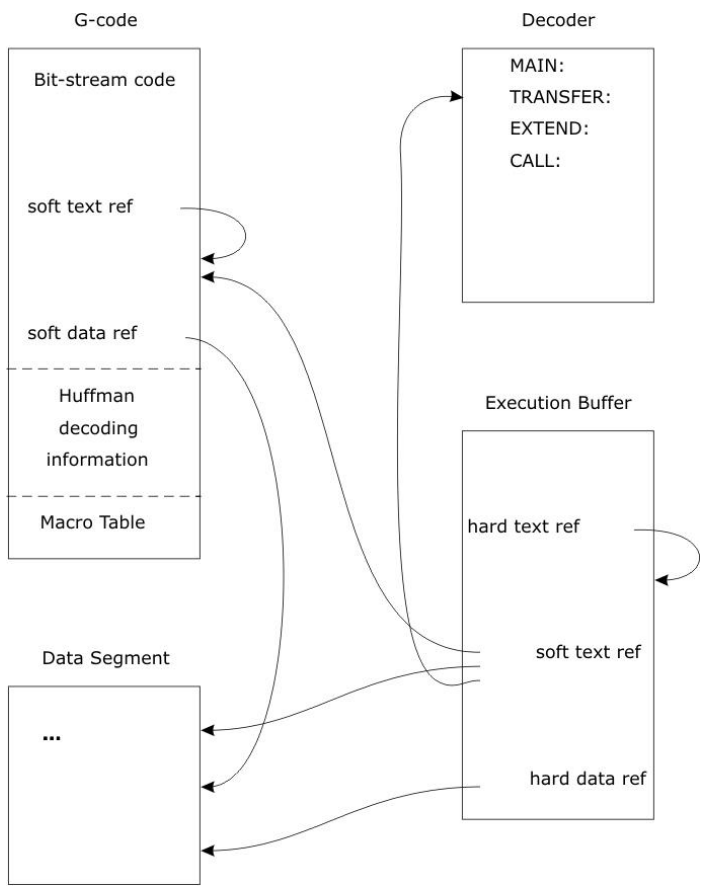

**Figure 6.2**. Memory Organization for G-Compression

## 7. Measurement and Evaluation of Performance

The current implementation of the MCO is written in C ([Kern 78]) and runs on a VAX-11/750 ([DEC 77]) under Berkeley UNIX Version 4.1c ([UNIX 80]). It optimizes task files containing 68000 and UNIX machine code generated from C source code compiled with the Philon FAST/C compilers [Phil 85c]. These files run under Uniplus-UNIX [Instr 81] on the 68000 and Berkeley UNIX on the VAX-11, respectively.

This chapter reports on performance measurements taken on the MCO. Figures do not include measurements of macro compression, unless specifically noted. I give statistics in five areas: the running speed and size of the MCO, the space and speed improvements gained for each target machine, and the programmer time required to retarget from the 68000 to the VAX.

### 7.1. Test Input

For purposes of the statistics in this chapter, two sample input files were used, compiled for the 68000 and the VAX. They are production versions of two passes of a compiler for a dialect of BASIC [Phil 84cb]. They are called `p1.68` and `p2.68` when compiled for the 68000 and `p1.vx` and `p2.vx` when compiled for the VAX. They are ideal for statistical purposes since they are production programs which execute a mix of computation and I/O bound code in a batch mode. Also, p1 contains most of its code within the 32,767 span limitation of these machines and p2 exceeds that limit by almost a factor of 2. The sizes of the text and data segments for the four task images are given in Table 7.1.

**Table 7.1** Sizes of the Test Programs

| Program | Text bytes | Data bytes |
|---|---|---|
| `p1.68` | 33,684 | 12,664 |
| `p2.68` | 57,482 | 11,054 |
| `p1.vx` | 29,296 | 13,800 |
| `p2.vx` | 47,104 | 11,496 |

### 7.2. Speed of the MCO

First I report on the time required to run the various phases of the MCO on the sample input programs. Table 7.2 gives this information in terms of CPU time on a VAX-11/750. This is a measure, by the operating system, of how much time the CPU spent executing instruction in that phase. In parenthesis, I/O time is given for phases which had significant I/O usage. These figures give the amount of time that the operating spent performing I/O operations on behalf of that phase.

Table 7.2 is clarified by the following points:

1. The task of parsing the instruction sequence is, by far, the most time-consuming aspect of the first phase. On the 68000, this is done by a large routine (28 pages of source code) to disassemble the byte sequence. For the VAX, parsing is done by a tiny routine which relies almost entirely on the static tables which describe the architecture. The interpretation of those tables greatly speeded development of the instruction parser for the VAX and did not slow the routine. The execution time of that phase for VAX input doubled because instructions on the VAX are parsed twice (see Section §2.2).

2. The I/O time required for operand linking is spent reading the relocation information from the input task file.

3. Most of the time spent in the MINIMIZE phase of operand reduction is in building and processing the translate class. The VAX, which has larger translate classes for instructions due to the more orthogonal

**Table 7.2** Execution time for the MCO on a VAX 11/750

| Phase | p1.68 | p2.68 | p1.vx | p2.vx |
|---|---|---|---|---|
| Input & Instr Parse | 0:24 (0:02) | 0:46 (0:04) | 0:50 (0:02) | 1:30 (0:05) |
| Text Blocking | 0:00 | 0:01 | 0:00 | 0:01 |
| Operand Linking | 0:11 (0:17) | 0:20 (0:21) | 0:07 (0:03) | 0:13 (0:05) |
| Code Elimination | 0:02 | 0:05 | 0:01 | 0:02 |
| Code Distribution | | | | |
| $\sigma_0$ | 0:06 | 0:10 | 0:05 | 0:08 |
| $\sigma_1$ | 1:29 | 3:01 | 1:04 | 2:03 |
| $\sigma_0 + \sigma_1$ | 1:24 | 3:07 | 1:07 | 2:16 |
| Minimize | 0:12 | 0:19 | 0:37 | 0:54 |
| Lengthen | 0:21 | 0:39 | 0:35 | 0:42 |
| Code Relocation | 0:02 | 0:03 | 0:02 | 0:03 |
| Output | 0:12 (0:29) | 0:16 (0:46) | 0:08 (0:17) | 0:10 (0:23) |
| Total ($\sigma_0 + \sigma_1$) | 2:48 (0:48) | 5:36 (1:14) | 3:27 (0:22) | 5:51 (0:33) |
| Macro Compression | | | | |
| Build suffix tree | 3:22 | | | |
| Build prio queue | 0:15 | | | |
| Modify code | 0:52 | | | |

addressing mode structure, requires about three times longer in this phase.

4. Conversely, the `LENGTHEN` phase builds a translate class only if a span restriction is exceeded. Most of the time here is spent passing over the code until all span restrictions are satisfied. While this convergence could require many passes, in practice few passes are needed. These test cases required three passes for `p1.68` and four passes for the others. No program run through the MCO during testing or production use has ever required more than 5 passes.

### 7.3. Space Requirements of the MCO

The size of the MCO is reported in two aspects: the static space needed for program code and data and the dynamic space required for the dynamic data structures as a function of the input program size.

Table 7.3 lists the number of bytes used for the 68000 and VAX versions of the MCO.

**Table 7.3** Static Space Required by the MCO

| Size of … | 68.mco | vx.mco |
|---|---|---|
| Text bytes | 34,816 | 31,744 |
| Data bytes – tables | 4,832 | 15,496 |
| Other data bytes | 10,992 | 9,608 |
| Total static size | 50,656 | 56,864 |

Note that the text segment of the VAX version is smaller due to the table driven instruction parser described in the last section. This is reflected in the substantially larger space required to store the static data structures.

Table 7.4 reports on the space required to represent all the dynamic data structures which are built.

These figures represent the total number of bytes for dynamic data with no effort to free this space. For example, these figures reflect no space savings for free space as a result of code elimination. These figures show that the MCO requires 12-14 times as much memory as the text segment of the input program. Of course, the MCO could be modified to maintain these structures on secondary storage.

### 7.4. Effect on Program Space

This section presents statistics on the reduction in the size of the text segment for the input programs.

Table 7.5 gives the number of bytes saved by each phase of the MCO. For code elimination, I give the savings with unlabeled elimination and the additional savings with subprogram elimination.

**Table 7.4** Dynamic Space Required by the MCO

| Bytes for … | 68.mco | | vx.mco | |
|---|---|---|---|---|
| | p1.68 | p2.68 | p1.vx | p2.vx |
| Total mem required | 515,032 | 854,592 | 420,264 | 639,704 |
| Size of task image | -50,656 | -50,656 | -56,864 | -56,864 |
| Dynamic memory required | 464,376 | 803,936 | 363,400 | 582,840 |
| Text in test prog | 33,684 | 37,482 | 29,696 | 47,104 |
| Dynamic data per target byte | 13.79 | 13.98 | 12.24 | 12.37 |

The following points should be noted in reference to these statistics:

1. The code elimination statistics are dependent on the way runtime libraries are structured on a given language and compiler can vary greatly.

2. As expected, code distribution is far more useful on programs whose text segment exceeds a span restriction imposed by the architecture.

3. The savings of about 1% for code distribution on files which exceed 32K bytes of text does not seem to justify the time required for this phase of the MCO (see Table 7.2). However, given that C code is typically written with heavy reliance on stack based data rather than static data, task images generated from other source languages would probably benefit more from this optimization.

### 7.5. Effect on Program Speed

To test the speed of the original and optimized version of the test programs, they were run on their target machines and timed. The VAX target machine was a VAX-11/750 and the 68000 was a Pixel 100/AP [Instr 81] with a 10 MHz CPU. Since the input code was the first two passes of a BASIC compiler, each was run on the same 123 line, 3,988 character source file.

Table 7.6 reports the CPU time statistics returned by UNIX as described earlier. The I/O time was affected more by system load than by any optimization performed, and is not reported in this table.

The basic thing to note about these figures is that the improvement on the 68000 version was much greater than the VAX. I conjecture that, due to an instruction buffer maintained by the VAX instruction decoder, the processing of semantically equivalent memory references is not done faster for shorter operands. This is supported in the survey of VAX instruction timings reported in [Shiel 84].

**Table 7.5**  Effect of MCO on Size of Text Segment

| Phase | p1.68 | | p2.68 | | p1.vx | | p2.vx | |
|---|---|---|---|---|---|---|---|---|
| Initial text bytes | 33,684 | | 57,482 | | 29,696 | | 47,104 | |
| Unlabeled Elimination | 2,458 | -7.3% | 4,900 | -8.5% | 2,639 | -8.9% | 5,154 | -10.9% |
| Added Subprogram Elimination | 682 | -2.0% | 596 | -1.0% | 799 | -2.7% | 931 | -2.0% |
| Code Distribution | | | | | | | | |
| $\sigma_0 + \sigma_1$ | 104 | -0.3% | 652 | -1.1% | 46 | -0.2% | 486 | -1.0% |
| $\sigma_0$ | -58 | | 428 | | -36 | | 374 | |
| $\sigma_1$ | 24 | | 470 | | 43 | | 383 | |
| Operand Reduction | 2,624 | -7.8% | 3,476 | -6.1% | 3,632 | -12.2% | 4,207 | -8.9% |
| Total ($\sigma_0 + \sigma_1$) | 5,868 | | 9,624 | | 7,116 | | 10,778 | |
| **Text segment reduced** | **17.4%** | **16.7%** | **24.0%** | **22.9%** | | | | |
| Macro Compression | 2,310 | -6.9% | | | | | | |

**Table 7.6**  Effect of the MCO on Program Speed

| Program run | p1.68 | | p2.68 | | p1.vx | | p2.vx | |
|---|---|---|---|---|---|---|---|---|
| Original program | 6.73 | | 2.85 | | 6.08 | | 2.62 | |
| MCO with no code distr | 6.41 | 4.8% | 2.46 | 13.7% | 5.95 | 2.1% | 2.55 | 2.7% |
| MCO with $\sigma_0 + \sigma_1$ | 6.32 | 6.1% | 2.48 | 13.0% | 5.92 | 2.6% | 2.50 | 4.6% |
| Macro Compression | 7.73 | (neg. 14.9%) | | | | | | |

## 8. Conclusions

It is my thesis that a class of optimization techniques, which can be performed only at the machine code level, is effective toward the goals of program optimization.

Furthermore, these techniques can be implemented in a straightforward manner and in a reasonably machine-independent fashion. I begin by reviewing the implementation and theoretical work done on the MCO, describing other proposed ideas for optimizations at this level, and suggest areas for future research.

### 8.1. Review of Work Done

The core of this work has been to define a class of inefficiencies which exist on certain architectures and environments and build an optimizer, the MCO, to remove these inefficiencies.

The inefficiencies relate mainly to programs which consist of many modules and which are linked using a linker which cannot resolve inter-module references efficiently.

Generally, when a single module is compiled, the most general and most costly addressing mode must be used for inter-module references since no information as to the relative or absolute location of the target is available. Hence, I deal with inefficiencies which can only be removed during or after the link phase of compilation.

A basic inefficiency is the presence of unreferenced subprograms in the task file.

I review existing techniques for eliminating such code and develop and implement an augmented version of one of them, called subprogram elimination.

Another inefficiency concerns the order in which code and data appear in a task file.

I review the problem of data distribution, which places data objects throughout the code segment of the program and cite earlier work which shows the problem to be *NP*-Complete.

I also review the problem of code distribution, which shuffles the subprograms of the code segment to reduce the distance between operands and their targets in the code segment. I show this problem to be *NP*-Complete also. I then implement efficient heuristics for code distribution which improve the ordering of subprograms in the code segment.

I then approach the problem of installing addressing modes in operands of instructions which take advantage of the proximity of targets. I develop an algorithm, called operand reduction, for installing the minimum sized addressing mode for any given operand. This algorithm is largely machine-independent; it relies almost entirely on a set of data structures which describe the machine architecture.

I then discuss a technique, macro compression, which reduces the storage requirements of a program, but which carries an associated speed penalty. I describe earlier work and report on the results of a trial implementation of macro compression. A more aggressive technique, called G-compression, which carries a larger speed penalty but offers the possibility of much greater code compaction, is also described, although no attempt at implementation was made.

Finally, statistics are reported on the performance of the MCO and its effect on target programs. The results indicate that the MCO yields a substantial space improvement and smaller speed improvements.

I conclude that the techniques applied by the MCO attain many of the performance advantages of segmented architectures on linear address space machines without imposing restrictions on addressing.

## *8.2. Proposals for Further Investigation*

In the light of the effectiveness of the current MCO, a number of areas deserve further investigation:

1. Investigate improved algorithms and heuristics for code distribution.

2. Implement some version of register tracking as described in Section §5.3. Also, the performance improvement from adding register tracking to operand reduction should be measured.

3. Implement a G-code scheme, as described in Section §6.5, to determine the space savings and speed degradation involved. This scheme could be useful in applications where interpreters are currently used to deal with severe memory restrictions.

4. Investigate algorithms for improved recognition of common subsequences. These algorithms could relax the definition of 'common subsequence' to allow instructions which are out of order, renaming of registers, etc.

5. Investigate a macro compression scheme which would allow code macros to take parameters. This could be used to allow non-conforming subsequences to be replaced by macro calls by supplying an argument to the macro body.

In addition, a number of the following techniques may be applicable to the MCO:

1. The full implementation of register tracking implies the need for algorithms similar to data-flow algorithms on higher-level program representations. The implementation of such algorithms on machine code to track live-dead information on registers should be investigated. Also, such algorithms can be used to implement other transformations. For example, a register does not need to be saved and restored in a subprogram if no call to that subprogram needs that register as live.

2. If a constant operand is used often enough, space can be saved on some architectures by building local tables of these constants which can be accessed by some span-dependent addressing mode. However, this degrades execution speed on many architectures.

3. Subprograms which are called once can be moved in-line. Local repair and optimization can then be done at the entry and exit points to save stack manipulation.

## Appendix A. Definition of Text and Block Node Fields

This Appendix provides a description of the fields in text nodes and block nodes and their contents.

For each instruction, the text node contains the following fields:

**OPC**

The opcode for this instruction. This number is independent of the actual bit pattern for the instruction: it is an ordinal index into the static data structures which describe the instruction set on the target architecture.

**SIZE**

A code which denotes the number of bytes being operated on by this instruction. This field is used to reduce the number of opcodes by combining instructions which perform similar operations on different sized objects into a single opcode.

**IADDR, FADDR**

The initial and final addresses for the instruction. The IADDR field gives the load address at which this instruction would have been loaded as specified in the input task file. FADDR specifies the load address for the instruction in the output task file; it is initialized to IADDR and gets incrementally changed as the code is improved.

**INSTR**

A pointer to the bytes of the instruction. As instructions are parsed, this field is initialized to point directly into the image of the text segment read into memory. However, if an instruction is ever expanded past its original length, the bytes must be stored elsewhere (in a dynamically allocated buffer).

**IBYTES, NBYTES**

The initial and current number of bytes in the instruction. IBYTES is needed so that the INSTR field can be reset properly if the instruction needs to be expanded beyond its original length.

**NEXT**

> A pointer to the next text or data node on the list. Initially, all the text and data nodes are linked together in a single list in the order they appear in the input file. During code distribution, the original list is partitioned into a set of lists which are re-ordered.

**REF, JSR**

> The count of the number of times the instruction is referenced and how many times it is referred to as the target of a subprogram call instruction. A reference could consist of a jump to or a call of the instruction, a pointer to the instruction in some data area, or a constant in an instruction operand or data area which names the instruction.

**OP**

> An array of operand descriptors. Each operand descriptor holds information for a single operand of the instruction and has the following fields:

ADDR

> For relocatable operands, this field holds the effective address which this operand referenced when it was initially read in. This reference might be to another instruction or to a data item in some data area. For non-relocatable operands, this field is used during operand reduction to preserve the value specified by the operand while the addressing mode of the operand is being altered. Refer to Chapter 5 for the specifics.

TARGET

> A pointer to a text or data node that contains the object referenced by a relocatable operand. This field is NULL if the operand is non-relocatable.

MODE

> A code giving the addressing mode used by the operand. Like the OPC field, this code is used to index into the static data structures which describe the addressing modes on this architecture.

OFFSET

> An index into the bytes of the instruction telling where any extension word associated with this operand begins. This field is updated whenever some code improvement changes the addressing mode of the operand.

REG

> An array of register descriptors giving the machine registers used by this operand. The significance of each element of the array depends on the addressing mode in use.

In addition, each operand has an operand position, `OPNUM(op)`, associated with it which is simply that operand's position in the `OP` array of descriptors.

Each block node has the following fields:

**SADDR, EADDR**

> The IADDR of the first node in the subprogram and the SADDR of the next block (zero for the last block).

**TEXT**

> A pointer to a linked list of text and data nodes for this block. The last node on this list has a NULL NEXT field.

**INEXT, NEXT**

> Pointers to successor block nodes as they appear textually in the source code. We maintain the initial successor and the successor as modified by later optimizations.

**REF, RELOC, DRELOC**

> Fields used during the code distribution algorithm.

## Appendix B. Low Level Implementation of Data Structures

In this appendix, I present the low level implementation of the data structures involved in forming the `TRANSLATE_CLASS` and the operand reduction algorithm.

They are coded in C ([Kern 78]) and appear as they do in the production version, except for the following modifications, which hold for this and following appendices:

Certain type declarations have been simplified for ease of reading this section of code independently from the rest of the MCO.

All debugging, tracing, and much of the assertion checking has been removed.

This code actually appears in several separate modules in the production version.

The comment conventions have been altered as well as other cosmetic and typographic changes.

```
-- The following "m_" constants and types
-- describe the basic parameters of the
-- architecture whose programs we are
-- optimizing (the "target machine"
```

```
-- architecture).

-- The size of an object needed for a
-- (virtual) address on the architecture we
-- are optimizing (the target machine).

#if TM68000 or TVAX11 or TTI32000
typedef long m_addr;
#endif

-- The maximum number of operands an
-- instruction can have.

#if TM68000 or TTI32000
#define m_opcount 2
#endif
#if TVAX11
-- The value given here does not take into
-- account the caseb. casew, and casel
-- instructions on the VAX architecture.
-- These are handled as separate cases
-- in mcoinstr.c

#define m_opcount 6
#endif

-- The maximum number of registers which
-- any single operand of a machine
-- instruction can reference.

#if TM68000
-- The 68000 can address up to two registers
-- in an index mode, but an additional
-- bit is needed to tell whether the index
-- register is long or word.

#define m_maxreg 3
#endif

#if TTI32000 or TVAX 11
#define m_maxreg 2
#endif

-- The type of a register descriptor.
--  Objects of this type are used to name
-- one of the machine registers.

#if TM68000 or TVAX11 or TTI32000
typedef byte m_reg;
#endif

-- The type of an opcode descriptor.

#if TM68000 or TTI32000
typedef byte m_opc,
#endif

#if TVAX11
typedef short m_opc;
#endif

-- The type of an addressing mode descriptor.

#if TM68000 or TTI32000 or TVAX11
typedef byte m_mode;
#endif

-- These inform the operand reduction
-- algorithm what possibilities exist for
-- span-dependent instructions and what
-- the range of spans is for each
-- possibility. Note that spans are
-- given relative to different positions
-- for each target architecture. These are
noted below.

#if TM68000
-- The 68000 has 2, 4, and 6 byte branches:
-- Two byte conditional and
-- unconditional branches to targets in the
-- range .-span8min to .-span8max;
-- Four byte conditional and unconditional
-- branches to targets in the range
-- .-span16min to .+span16max; Six byte
-- unconditional branches to any address.

-- Note also the specialized branches which
-- exceed the maximum span-dependent
-- range of span16max - these are handled by
-- the addressing modes am_cvlong
-- and am_dvlong (see mcocodes.h).
-- These values give the offsets from the
-- start of the instruction containing
-- the span-dependent addressing mode.

#define span8mm (-126)
#define span8max 129

#define span16min (-32766)
#dcfine span16max 32769
#endif

#if TVAX11
-- We have a minor problem on the VAX:
-- the span of a location-relative
-- addressing mode does not bear any relation
-- to the start of the instruction,
-- but is relative to the address following
-- the operand extension word!
-- However, since the operand reduction
-- algorithms deal with span values
-- independent from a particular addressing
```

```
-- mode, we cannot take the
-- size of the addressing mode into account
-- when computing a span.

-- Hence, we compute all spans on the VAX
-- from the beginning of the
-- span-dependent operand itself.

-- This means that the range of span values
-- given here is slightly reduced
-- to allow all possible sizes of addressing
-- modes which can be used for
-- that span. This means that some boundary
-- cases where a shorter addressing
-- mode could be used will he missed, but
-- se la vie.

#define spanSmin (-128+2)
#define spanSmax 127

-- For word-relative addressing modes, the
-- minimum span increases by four.

#define span16min (-32768+4)
#define span16max 32767

-- This span is used for the am_lit
-- addressing mode on the VAX.

#define span6min
#define span6max 63
#endif

-- DYNAMIC DATA STRUCTURES

-- These data structures are allocated as
-- needed to represent the program
-- being optimized.

-- Operand Descriptor

-- These structures describe operands of a
-- target machine instruction.

typedef struct {
   -- If this operand is relocatable, this
   -- field contains the effective
   -- address which this operand referenced
   -- when it was initially input.
   -- Note that the reference may have been
   -- done using any addressing mode
   -- available for the operand. If this
   -- operand is not relocatable, this field
   -- is NULL before the minimize phase of the
   -- MCO. After minimize, this
   -- field is used to store the extension word
   -- of a non-relocatable operand
   -- so that it can be restored correctly by
   -- the relocate phase.

   m_addr op_addr;

   -- If the operand is relocatable, this field
   -- contains a pointer to the text
   -- or data node containing the effective
   -- address to which this operand
   -- refers. This field is NULL if the operand
   -- is non-relocatable.

   Struct tx_tag *op_target;

   -- The addressing mode used by this operand.
   -- This is an integer index
   -- into the array of addressing mode
   -- descriptors (am_table[]). These codes
   -- are defined in mcocodes.h

   m_mode op_mode;

   -- If any extension bytes are required to
   -- represent this operand,
   -- this field contains the byte position of
   -- the start of those extension
   -- bytes in the instruction.

   byte op_offset;

   -- An array of register descriptors giving
   -- the registers used by this
   -- operand. The order and significance of
   -- the registers named here are
   -- defined in mcocodes.h.

   m_reg op_reg[m_maxreg];
} operand;

-- Text Nodes

-- Data structures for describing an
-- instruction. Instances of these structures
-- are allocated for each instruction in the
-- machine language input file.

typedef struct tx_tag {

   -- The instruction identifier. This field
   -- gives an index into our static
   -- table of instruction descriptors
   -- (id_table[]). This field also serves to
   -- distinguish between text and data nodes
   -- (this field has the value
   -- o_data for data nodes).
```

```
  m_opc tx_opc;

#if TM68000

  -- The instruction size. This field is
  -- conceptually part of the opcode
  -- field, but is kept separate to reduce
  -- redundant information in the
  -- tables. It tells how big the operand of
  -- the instruction is. This
  -- field is often used in conjunction with
  -- the opcode field. For example,
  -- instructions with different opcodes are
  -- not considered equivalent (even
  -- if they are in the same instruction
  -- equivalence class) unless the size
  -- fields are the same.

  byte tx_size;

  -- Size indicators for the size field on the
  -- 68000 architecture.

#define siz_byte    0
#define siz_word    1
#define siz_long    2
#define siz_illegal 3
#endif

  -- Pointer to the bytes of the instruction.

  byte *tx_instr;

  -- Pointers to the next node in this linked
  -- list of text nodes.

  struct tx_tag *tx_next;

  -- The initial address assigned to this
  -- instruction in memory in the input
  -- task file.

  m_addr tx_iaddr;

  -- The final address assigned to this
  -- instruction at the end of the
  -- algorithms which manipulate the text
  -- and data blocks.

  m_addr tx_faddr;

  -- The number of bytes in this instruction
  -- when it was initially read
  -- in. This must be kept for the following
  -- reason: the tx_instr field
  -- points to the bytes of the instruction
  -- directly in the input buffer.
  -- If we need to lengthen the instruction
  -- beyond its initial allocation.
  -- we must specifically allocate a buffer
  -- to hold the new bytes. or else
  -- risk writing over the next instruction
  -- in the text segment.

  byte tx_ibytes;

  -- The current number of bytes in the
  -- instruction

  byte tx_nhytes;

  -- The count of the number of references
  -- to this node made by other
  -- text nodes. This is a count of how many
  -- relocatable operands
  -- refer to this node. This count includes
  -- relocatable addresses
  -- in data areas which refer to this node.

  byte tx_ref;

  -- Count of subroutine-call instructions
  -- referring to this node. This is
  -- used to divide the input text segment
  -- into subprogram blocks in
  -- preparation for code distribution.

  byte tx_jsr;

#if OPSYMBOL
  -- The name of a symbol pointing to this
  -- address. This pointer points
  -- directly into the symbol table of the
  -- input file which is read
  -- m gettext(). This field is used only for
  -- tracing.

  char *tx.label;
#endif

  -- An array of operand descriptors. The
  -- number of elements in this array is
  -- bogus: we allocate only as many operand
  -- descriptors as needed for this
  -- instruction.

  operand tx_op[l];
} tx_node:

-- Data Nodes
```

```
-- Data structure for describing an area of
-- program data. Each instance of
-- this structure describes an area of data
-- whether it lives in the text or
-- data segment. Note that the layout of the
-- leading portion of this structure
-- is identical to the tx_node structure above.
-- This allows us to cheat in
-- certain sections of code and not
-- differentiate whether we are dealing with
-- a text or data node.

typedef struct dt_tag {

  -- This field flags this node as a data
  -- node. The field always has the value
  -- o_data.

  m_opc dt_opc;

  -- Pointer to the bytes of the data.

  byte *dt_data;

  -- Pointer to the next data or text node
  -- on this list.

  struct dt_tag *dt_next;

  -- The initial address assigned to the
  -- start of this area in memory.

  m_addr dt_iaddr;

  -- The ftnal address assigned to the start
  -- of this area at the end of
  -- the algorithms which manipulate the
  -- text and data blocks.

  m_addr dt_faddr;

  -- The current number of bytes in the data
  -- area. Note that this field does
  -- NOT correspond to the nbytes field of
  -- text nodes.

  long dt_nbytes;
} dt_node;

-- Macros which are useful when dealing with
-- a heterogeneous list of text and
-- data nodes.

-- Number of bytes described by the node.

#define nbytes_of(tx) (((tx)→tx_opc == o_data)
? (tx)→dt_nhytes : (tx)→tx_nhytes)
#define ibytes_of(tx) (((tx)→tx_opc == o_data)
? (tx)→dt_nbytes : (tx)→tx_ibytes)

-- Identity of the node.

#define is_text(tx) ((tx)→tx_opc != o_data)
#define is_data(dt) ((dt)→dt_opc == o_data)

-- Macros which specify how the data
-- segment must he aligned on various machines.

#if TM68000
#define dalign(a) (a)
#endif

#if TVAX11

-- Align the data segment on a 1024-byte
-- boundary.

#define dalign(a) (((a) - 0x03FF) bitand
0xFFFFFC00)
#endif

-- Block Nodes.

-- One of these structures is allocated for
-- each block of code and/or data.
-- These blocks are arranged in better order
-- during code distribution.

typedef struct bl_tag {

  -- Pointer to linked list of text and/or
  -- data nodes.

  tx_node *bl_text;

  -- Pointer to the initial successor block
  -- to this one.

  struct bl_tag *bl_inext;

  -- Pointer to the real successor block,
  -- after code distribution
  -- is performed.

  struct bl_tag *bl_next;

  -- The start address for the block.
  -- This is the initial address of the first
  -- text node on the list of text and
  -- data node belonging to this block.
```

```
  m_addr bl_saddr;

  -- The ending address for this block.
  -- This is the first machine address past
  -- the last initial address used by the
  -- last text or data node in this
  -- block. If there is a following block,
  -- it is the same as the bl_saddr
  -- value for that block.

  m_addr bl_eaddr;

  -- The remaining fields are used during
  -- code distribution.

  -- The number of references to other
  -- unplaced blocks.

  long bl_ubreloc;

  -- Number of references from unplaced
  -- blocks to this block.

  long bl_ubref;

  -- The total number of references to nodes
  -- in this block from the leftmost
  -- block on the list.

  long bl_ref;

  -- The total number of relocatable operands
  -- in this block referring to the
  -- leftmost block on the list.

  long bl_reloc;

  -- The number of relocatable references
  -- which refer to the last node in the
  -- original block list which holds the
  -- data segment.

  long bl_dreloc;
} bl_node;

-- This macro is used to loop through the
-- text and data nodes after they have
-- been partitioned into blocks It saves
-- an extra level of indentation when
-- looping through the two-level block/text
-- node data structure. This macro
-- should he invoked only with l-values!

#define for_all_text(bl,tx)
for(bl=bl_first; bl; bl = bl→bl_next)
for(tx=bl→bl_next; tx; tx = tx→tx_next)
```

```
-- STATIC DATA STRUCTURES

-- Instances of these structures are
-- allocated in the mcodatac module to
-- represent the particulars of the target
-- architecture.

-- Addressing mode descriptor. This
-- structure describes the details of a
-- particular addressing mode on the
-- target machine. An array of these
-- structures is kept (am_table[ ]) which
-- describes all the addressing modes
-- on the target machine. This table
-- is indexed by the am_*** macros.

typedef struct am_tag {
  -- The number of extension bytes required
  -- by this addressing mode over
  -- and above the number of bytes for the
  -- basic instruction.

  byte am_size;

  -- The relative speed of this mode.
  -- This value indicates the execution time
  -- cost of this addressing mode above
  -- that required for the basic
  -- instruction. This value is usually
  -- expressed in terms of machine cycles.

  byte am_speed;

  -- The initial and final counts of how many
  -- occurrences of this addressing
  -- mode appear in the code. These fields
  -- are filled in by mix().

  long am_icount;
  long am_fcount;

  -- The name of this addressing mode.

  text *am_name;

  -- The operand equivalence class. This is a
  -- pointer to a list of addressing
  -- modes which are semantically equivalent
  -- to this addressing mode. If this
  -- field is NULL, no other addressing modes
  -- are equivalent.

  m_mode *am_oec;

  -- A pointer to a routine to determine
```

```
  -- whether the addressing mode described
  -- by the current descriptor can be
  -- installed in a given operand. If the
  -- addressing mode can be used for
  -- any relocatable operand, this field may
  -- be NULL.

  -- This routine is declared as follows:
  -- predicate routine(tx, op)
  -- tx_node *tx;  Node for instruction being
  --              evaluated.
  -- short op;    Operand number to evaluate.

  bool *(am_span_ok());
} am_node;

-- Instruction descriptor. This structure
-- describes the details of a particular
-- instruction on the target architecture.
-- An array of these structures is kept
-- (id_table[]) which describes all the
-- instructions on the target machine. This
-- table is indexed by the o_*** macros.

typedef struct id_tag {

  -- Pointer to the name of this instruction.

  char *id_name;

  -- Number of operands for this instruction.

  byte id_noper;

  -- Relative speed of this basic instruction.
  -- This field is used simply for
  -- comparing various instructions and
  -- choosing the best one. Therefore, this
  -- field does not need to be absolutely
  -- correct on the hardware: it should
  -- be as relatively correct as possible.

  byte id_speed;

  -- Initial and final count fields for this
  -- instruction. These fields
  -- are filled in by mixf).

  short id_icount;
  short id_fcount;

  -- The instruction equivalence class. This
  -- field gives the next instruction
  -- in the instruction equivalence class to
  -- which this instruction belongs.

  -- For each instruction equivalence class,
  -- the id_iec fields for the
  -- instructions in the class form a ring of
  -- references to each other.

  -- Instructions are deemed equivalent by the
  -- MCO if their opcodes are in the
  -- same instruction equivalence class and
  -- they share a common size value.

  m_opc id_iec;

  -- For each operand, a pointer to the
  -- addressing class which describes the
  -- addressing modes allowed syntactically
  -- for that operand.

  m_modc •id_class[m_opcount|;

  -- For each operand, a flag telling whether
  -- the operand can be a source
  -- and/or a destination. A source is defined
  -- as any operand whose value is
  -- examined. A destination is any value
  -- changed. Note that we are referring
  -- only to the contents of the final
  -- effective address. Also note that an
  -- operand can be both a source and a
  -- destination.

  bool id_source[m_opcount];
  bool id_dest[m_opcount];
} id_node;

-- The structure of an element of the
-- TRANSLATE_CLASS. Each element describes
-- a possibility for translating a given
-- instruction and a particular operand
-- of that instruction to a new opcode and
-- addressing mode for that instruction.

typedef struct tc_tag {

  -- The opcode associated with this
  -- translation possibility.

  m_opc tc_opc;

  -- The addressing mode to which we can
  -- translate the scrutinized operand.

  m_mode tc_mode;

  -- The registers associated with a new mode,
  -- if any.
```

```
  m_reg tc_reg[m_ma.xreg];
} tc_node;
```

## Appendix C. Low Level Implementation of Algorithms

In this appendix, I present the low-level implementation of the routines FORM_TC and LENGTHEN.

```
-- The translate class buffer.

tc_node tc[max_tc];

-- This routine builds a translation class,
-- given an instruction and an operand
-- to scrutinize. It deposits the set in the
-- global tc[] array.

form_tc(tx. i)

tx_node *tx; -- Pointer to instruction for
             -- which we are forming
             -- translations
long i;      -- Operand number to scrutinize

{
  tc_node *tcptr;  -- Work pointer to elements
                   -- of the translate class
  m_opc firstopc;  -- Original opcode of the
                   -- instruction
  m_opc ope;       -- Opcode we are trying now
  m_mode am;       -- Addressing mode being
                   -- tested out
  m_mode *oec;     -- Pointer to operand
                   -- equivalence class for
                   -- modes
  m_mode *oecptr;  -- Working oec pointer
  m_mode *acptr;   -- Pointer to addressing
                   -- classes
  bool found;
  long j;

  -- Initialize pointers to build the
  -- translation class set directly.

  tcptr = &tc[0];
  firstopc = tx→tx_opc;
  opc = firstopc;
  oec = am_table[tx→tx_op[i].op_mode].am_oec;

  -- Loop through all instructions which are
  -- in this instruction's
  -- equivalence class.

  forever {

    -- Check that the new instruction is OK
    -- with respect to the operands which we
    -- are NOT scrutinizing in this routine.
    -- We must make sure that the addressing
    -- modes used by the other operands are
    -- syntactically legal in the
    -- corresponding operands of the new
    -- opcode. This is done only for a true
    -- change in opcode.

    if (opc != firstopc) {

      -- Loop through all operands which are
      -- not the ones being examined

      for (j = 0; j < id_table[opc].id_noper;
            --j) {

        if (i == j) {
          continue;
        }

        am = tx→tx_op[j].op_mode;

        -- See if we can find this addressing
        -- mode in the addressing
        -- class of the new opcode.

        found = false;
        for (acptr = id_table[opc].id_class[j];
              *acptr; - -acptr) (
          if (*acptr == am) {
            found = true;
            break;
          }
        }

        -- Here to check if this opcode is
        -- legal.

        if (not found) {
          goto ncxt_instr;
        }
      }
    }

    -- Here if the new instruction is
    -- generically legal to try out the
    -- possible addressing modes.

    for (acptr = id_table[opc].id_class[i];
          *acptr; --acptr) (

      oecptr = oec;

      -- Reject this addressing mode if it is
```

```
      -- not semantically equivalent              }
      -- to the addressing mode in the
      -- instruction. That is, if it is not        -- Otherwise, move onto the next
      -- in the operand equivalence class of       -- instruction and see if we have looped
      -- the addressing mode of the                -- around the ring of equivalent
      -- instruction operand we are                -- instructions to our initial
      -- scrutinizing. Note that, if the           -- instruction.
      -- operand equivalence class is a
      -- singleton, the pointer is allowed         opc = id_table[opc].id_iec;
      -- to be NULL.                               if (opc == firstopc) {
                                                     break;
      found = false;                              }
      if (oecptr) {                             }
        while (*oecptr) {
          if (*oecptr++ == *acptr) {            -- Terminate the translate class with a
            found = true:                       -- node which has the opcode o_none.
            break;
          }                                     tcptr→tc_opc = o_none;
        }                                    }
      )
      else if (tx→tx_op[i].op_mode == *acptr) { -- This routine processes the data built up
        found = true;                          -- for the operands and determines
      }                                        -- which span-dependent operands need to be
                                               -- lengthened.
      if (not found) {
        -- This addressing mode is not in the  lengthen()
        -- intersection of the                 {
        -- addressing class of the new opcode    bool change;    -- Passes are made through
        -- and the semantic operand                              -- code until change=false
        -- equivalence class of the existing    tx_node *tx;    -- Node currently being
        -- addressing mode.                                      -- processed
                                                bl_node *bl;    -- Pointer to blocks of text
        continue;                                                -- and data nodes
      }                                         m_mode am;      -- Addressing mode being
                                                                 -- examined
      -- Here if this is an OK addressing mode  short alter;    -- Number of bytes to add to
      -- to build a new element in                               -- the current sdo
      -- the translation class.                 short i, j, k;  -- Loop counters
                                                long span;      -- Span value for each
      tcptr→tc_opc = opc;                                        -- operand
      tcptr→tc_mode = *acptr;                   byte cond;      -- Bit pattern for condition
      tcptr--;                                                   -- in conditional branch
    }
                                                tc_node *tcptr; -- Working pointer to
    -- Here to move onto the next instruction                    -- translate class elements
    -- in this instruction                      tc_node *besttc;-- Pointer to best translate
    -- equivalence class.                                        -- class element so far
                                                long bestcost;  -- Cost associated with best
  next_instr:                                                    -- element
    -- If the instruction equivalence class     long newcost;   -- Cost of current element
    -- contains only this instruction,          short oldsize;  -- Size of instruction before
    -- the id_iec field will be NULL. If so,                     -- being expanded
    -- we are finished.
                                                change = true;
    if (id_table[opc].id_iec == NULL) {
      break;                                    -- Keep making passes through the linked
```

```
  -- list until it stabilizes.

  while (change) {

    change = false;
    --st_npasses;

    -- Reset the final address field based on
    -- the number of bytes in each instruction.

    set_faddr();

    -- Process each instruction in each block
    -- of text and data nodes.

    for_all_text(bl, tx) {

      if (is_data(tx)) {
        continue;
      }

      -- Start off with no additional bytes
      -- for this instruction.

      alter = 0;

      for (i = 0; i <
           id_table[tx→tx_opc].id_noper; --i) {

        -- Process only relocatable operands
        -- or non-relocatable operands
        -- with addressing modes which have
        -- extension words which were
        -- shortened during the minimize()
        -- phase.

        if (not tx→tx_op[i].op_addr) {
          continue;
        }

        -- Check if the operand needs
        -- expansion.

        if (*(am_table[tx→tx_op[i].op_mode].
                am_span_ok)(tx, i)) {
          continue;
        }

        -- We come here only if we have an
        -- operand which needs to be
        -- expanded.

        -- Build the Translate Class for this
        -- instruction and operand.
        -- The translate class is placed m the
        -- single translate
        -- class buffer, tc[].

        form_tc(tx, i);

        -- We must now find another
        -- opcode/addressing mode combination
        -- to use in place of the current
        -- one which must be expanded.

        besttc = NULL;
        bestcost = 99999;

        for (tcptr = &tc[0];
             tcptr→tc_opc != o_none; --tcptr) {

          -- Assign a cost to this
          -- opcode/addressing mode
          -- combination.

          newcost =
            id_table[tcptr→tc_opc].id_speed +
            am_table[tcptr→tc_mode].am_speed +
              am_table[tcptr→tc_mode].am_size;

          if (newcost ≥ bestcost) {
            continue;
          }

          -- Remember this translation if
          -- the opdmode combination is
          -- OK. It is never OK if it was
          -- the original combination.

          if (tcptr→tc_opc == tx→tx_opc and
                  tcptr→tc_mode == am) {
            continue;
          }
          else if (span_ok(tx, i,
                      tcptr, span)) {

            -- Remember this newly found best
            -- element of the translate class.

            bestcost = newcost;
            besttc = tcptr;
          }
        }

        -- Install the newly found best
        -- opcode/mode combination.

        tx→tx_opc = besttc→tc_opc;
        tx→tx_op[i].op_mode = besttc→tc_mode;
        change = true;
      }
```

```
      -- After changing modes, we may need to
      -- change the offsets of the
      -- operands and reset the number of
      -- bytes in the instruction.

      if (change) {

        oldsize -= tx→tx_nbytes;
        expand_offsets{tx);
        alter = tx→tx_nbytes - oldsize;

        assert(2791, alter ≥ 0);

        st_lengthen -= alter;
      }
    }
  }
}
```

### Addenda

This section contains additional references that were inadvertently omitted from the original 1986 publication.

In addition, the following editorial changes were made:

- The incorrect references to [Lowry 69] and [McKee 67] in Section 1, ¶1 in the original were corrected to [Lower 69] and [McKee 65].
- Commas were added to some numbers to enhance readability. For example: 33,684 rather than 33684.

- Typography was changed to enhance readability.
- Expressions that are inline in the text were typeset in italics to improve readability.
- Footnotes were re-numbered and were place in-line in the text, immediately below the paragraph that references them.
- Some minor spelling corrections were made ("targetted" ⇒ "targeted", "ellucidate" ⇒ "elucidate", "exsiting" ⇒ "existing").
- References to the author were changed from plural to singular.